\documentclass[12pt,a4paper]{article}
\usepackage{amsmath}
\usepackage{amsthm}
\usepackage{longtable}
\usepackage{graphicx}
\usepackage{enumerate}
\usepackage{natbib}  
\usepackage{url}

\newcommand{\blind}{0}

\begin{document}

\def\spacingset#1{\renewcommand{\baselinestretch}%
{#1}\small\normalsize} \spacingset{1}

\if0\blind
{
  \title{\bf A Structural Causal Model for
  Electronic Device Reliability: From  Effects to Counterfactuals
   }
  \author{Federico Mattia Stefanini\hspace{.1cm}\\
    Department of Environmental Science and Policy, 
    University of Milan, \\
    Via Celoria 2, 20133 Milan, Italy\\
    federico.stefanini@unimi.it \vspace{0.5cm}\\
    Nedka Dechkova Nikiforova\\
    Department of Statistics Computer Science Applications "G. Parenti", \\
    University of Florence,
    viale Morgagni 59, Florence, 50134 Florence, Italy\\
    n.nikiforova@unifi.it\\
    and \hspace{.1cm}\\
    Rossella Berni\thanks{
    The authors gratefully acknowledge  the financial support
    from the Italian Ministry of University and Research (MUR) under the Project of Relevant National Interest (PRIN 2022), project code E3DM - CUP master:B53C24006390006; CUP: B53C24006400006.
    } \\
    Department of Statistics Computer Science Applications "G. Parenti",\\
    University of Florence, viale Morgagni 59, Florence, 50134 Florence, Italy\\
    rossella.berni@unifi.it}
  \maketitle
} \fi

\if1\blind
{
  \bigskip
  \bigskip
  \bigskip
  \begin{center}
    {\LARGE\bf A Structural Causal Model for  reliability }
\end{center}
  \medskip
} \fi

\bigskip
\begin{abstract}
Electronic devices exhibit changes in electrical resistance over time at varying rates,
depending on the configuration of certain components.
Since measuring overall electrical resistance requires partial disassembly,
only a limited number of measurements are performed over thousands of operating hours. 
This leads to censored failure times, whether under natural stress  or under accelerated stress conditions.
To address these challenges, including device-specific failure thresholds,
a parametric structural causal model is developed to  extract information from both observational and experimental data,  with the aim of estimating causal effects and counterfactuals, regardless of the applied stress regime. 
Synthetic data are used to illustrate the main findings.
\end{abstract}

\vfill

\spacingset{1}

\section{Introduction}
\label{sec:intro}

The estimate of reliability  \citep{Meeker2021} \citep{Singpurwalla2006}
can be expensive and time-consuming,
especially when dealing with electronic devices that are designed
to be highly reliable.
Modern reliability assessments  
combine    information from multiple sources
rather than relying   solely on  test data \citep{Hund2020}.
Observational data may come from end users, thus the analyst has
no control over the data-collection process.
The consequent biases that affect reliability estimates include:
(i) selection bias if the available data are not representative of the target
population because the data-collection process favors certain types of
devices or operating conditions;
(ii) confounding bias when omitted variables in the statistical model
affect both the attribute under investigation and the response variable,
confounding the relationship between them.

In this study, we consider the electrical resistance of electronic devices, focusing on factors such as surface finish, component type, and pin size.
To measure the electrical resistance of a device, the machine in which it is
installed must be partially disassembled, thus, besides the initial
measurement  at the beginning of the work in production, only three to five measurements
of electrical resistance are typically performed over a time window
of several thousand working hours.
Measurement times are also well spaced apart in randomized experiments performed 
in controlled conditions  to study the effect of one or more experimental factors,
like the surface finish and the size of pins.
Interval and right censoring follow in these cases
because   the exact failure time at which the electrical resistance increases
over the defined threshold is  not available.
It is worth noting that
even basic designed experiments
are often performed under an accelerated stress
regime, thus statistical models that assimilate data coming from different sources
must also accommodate  alternative regimes.

Most of the  peculiar features of the above-described context
may be faced by developing a causal model
able to extract information
from observational as well as  experimental data.
Different approaches to build a causal model have been developed
in the last thirty years  \citep{Berzuini2012},
thus we do not provide an exhaustive review of the  topic.
\citet{dominici} and \citet{brand} are two recent review papers in this field, including machine learning methods in causal inference. 
More precisely, \citet{brand} deal with theoretical and terminological aspects, as well as with applications strictly related to the sociological field. 
\citet{dominici} provide a detailed review of the potential outcomes
framework for causal inference, explicitly considering
the advantages and disadvantages of machine learning approaches to causal inference;
moreover, the authors highlight open problems that remain unaddressed by current machine learning techniques.
In \citet{brand},
a more comprehensive overview of the different approaches for causal inference is provided, with particular attention to path analysis and structural equation modeling in the sociological field.
More specific review papers are provided by \citet{degtiar} and \citet{Vonk2023}. To this end, \citet{degtiar} deal with methods for addressing external validity bias, by specifically concentrating on the approaches for generalizability and transportability.
In \citet{Vonk2023}, the most popular approaches 
in the causal inference literature are described, starting from  the set of  assumptions they 
rest on, and classified according to  
Pearl's Causal Hierarchy, i.e.,  seeing-doing-imagining,
\citep{Bareinboim2022}.

Structural Causal Models (SCMs, \citeauthor{Pearl2009}{, 2009}) 
are adopted here as the working framework because, in our context, they offer a cognitively appealing way to define direct causal relationships through deterministic functions and to qualitatively represent the set of considered relationships using causal Directed Acyclic Graphs (DAGs).
An introduction to the structural causal modeling framework  in reliability studies is
provided  by  \cite{Hund2020}, 
who deal with causal modeling tools in the field of reliability assurance for biased data sources. 
The authors consider two types of bias, i.e.,  selection bias and confounding bias, also performing a sensitivity analysis. 
In \citet{RuizTagle2022}, a novel probabilistic approach to counterfactual reasoning is introduced in the context of system safety. 
To this end, the authors propose a methodology specifically oriented toward system safety, employing Bayesian network models to evaluate counterfactual hypotheses.

In this work, we develop a SCM to investigate changes in electrical resistance as a function of electronic device configuration, relative humidity,
and operating time (in kilohours).
The proposed model provides Bayesian estimates such as the failure time of an electronic device and the counterfactual value of electrical resistance under a hypothetically different observation time.
Section \ref{sec:solderproc} describes the main features of the reliability context assumed in this study.
Section \ref{sec:scmDAGS} introduces the basic concepts of non-parametric SCMs and DAGs,
and presents a simple context involving 
only one measurement time for electrical resistance testing.
In Section \ref{sec:paraSCM}, a parametric SCM for observational studies is presented,
in which three measurement times are considered.
The structural equations and the probability density functions (after  marginalization) are defined and  describe the submodel tailored for studies conducted under accelerated stress conditions.
Section \ref{sec:causaestisynth} introduces
two synthetic datasets, one for each stress regime.
 The prior distributions on model parameters are then defined.
In this section, the  average difference in the  increase of resistance under varying configurations
is analyzed in the accelerated stress regime, and a reliability formula is proposed.
The no-stress regime is then examined and the predictive distribution of the failure time
under a given configuration of the electronic device is derived, together with useful
counterfactuals.
Section \ref{sec:conc} provides  final remarks and conclusions.

\section{Assessing the electrical resistance of electronic devices}
\label{sec:solderproc}
 
The electrical resistance  $y_{0,j}\in (0,\infty)$, measured in ohms, is recorded for  an electronic device $j$
prior to the start of its operation, at time  $w_{0,j}=0$ kilohours (i.e., thousands of hours).
In the following analysis, the index $j$ is consistently used to denote devices (i.e., statistical units), and $n$  indicates the sample size.
Other indices, such as $i$ and $r$, may refer to different objects depending on the specific formula or context in which they appear.
At  time $w_{1,j} >0$,  the electrical resistance
$y_{1,j}\in (0,\infty)$ is  measured again.
The random variable   $W_{f,j}$
refers to the (unknown) failure time
of   device $j$,
with sample space $\Omega_{W_f} = (0,\infty)$:
the failure  occurs 
if the electrical resistance, measured in ohm,
increases by  $10\%$, thus   reaching the value $1.1 ~ y_{0,j}$.

The reliability function, 
$R(w_{j}\mid \xi) = 1- F(w_{j} \mid \xi) =
 P[W_{f,j} > w_j \mid \xi]$ in context $\xi$,
also known as survival function,
describes the probability that a device is still 
 operational at time $w_j$.
By substitution, we obtain: 
\begin{align}\label{eq:reliability01}
    P[W_{f,j} > w_j \mid \xi] = P[Y_{1,j} < 1.1 \cdot y_{0,j}\mid W_{1,j} = w_j, y_{0,j}]
\end{align}
where the electrical resistance $Y_{1,j}$
may depend on selected experimental factors besides the elapsed
time $w_{1,j}=w_j$;
some of them may also have an effect on the initial response $Y_{0,j}$.

Life-testing  may be performed under different regimes \citep{Singpurwalla2006}:
(i) the  regular no-stress  (NS) condition where the electronic device works
in standard conditions;
(ii) the accelerated stress  (AS) condition where  the amount of  stress acting on the 
tested electrical device is elevated. 
In our context, acceleration is performed by applying thermal cycles where
the temperature   cyclically varies from
 $0~ [C^{\circ}]$ to $100~[C^{\circ}]$ along time in controlled experiments. 
The AS regime is often adopted  when the design of an experiment 
under the NS regime
would entail a large amount of test time before yielding failures.
Observational studies are typically performed
under the NS regime.
The regime variable $A_j$   indicates  whether
on device $j$ the AS regime, $A_j=a_2$, or the NS one, $A_j=a_1$,
is in force.

Variables $W_{i,j}$, with $i =1, 2, 3,\ldots$ represent the subsequent measurement times
for observation $j$, regardless of the operating regime (NS or AS), starting from the beginning of the study. 
The number of elapsed hours in the AS regime
is naturally mapped into the number of thermal cycles performed.

In the following analysis, three experimental factors describing key characteristics of an electronic device are considered.
For  device $j$, the type of  surface finish is indicated as $x_{S,j}$, 
and $\Omega_{x_S}=\{1,2,\ldots,n_S\}$ is the sampling space;
the type of component is $x_{T,j}$, with
$\Omega_{x_T}=\{1,2,\ldots,n_T\}$,
while for the number of pins  $x_{P,j}$ we have $n_P$ levels,
$\Omega_{x_P}=\{1,2,\ldots,n_P\}$.
Following the  literature, the change in electrical resistance
of a device over time also
depends on the humidity level during its operation.
The  relative humidity, $x_{H,j}$, 
is represented by $n_H$  integers,
each one corresponding to  an interval in a  partition defined over a scale
ranging from $0\%$ to $100\%$.
In  our context, the partition consists of two elements:
the normal range, $x_H = -1$, and high range,  $x_H = 1$.

\section{Structural Causal Models and the minimal DAG}
\label{sec:scmDAGS}

A Non-Parametric Structural Causal Model (NP-SCM, Pearl 2009, Definition 7.1.1, modified)
is a triple  
$\mathcal{S}= \langle  U,V,\mathcal{H} \rangle$, where 
$V=\{V_1,V_2,\ldots,V_i,\ldots\}$ is a finite collection of endogenous variables.
Each variable $V_i \in V$ is explicitly related to other
variables  in $V$ and  in $U=\{U_1,U_2,\ldots,U_i,\ldots\}$,
the finite set of exogenous-background variables.
The exogenous variable $U_i$ collects  all other causal determinants
of $V_i$ not explicitly included in the model
as endogenous variables.
The set  of deterministic functions
$\mathcal{H}=\{h_1,h_2,\ldots,h_{i}, \ldots\}$ 
contains one function for each endogenous variable;
for node $V_i$, the function  $h_i()$   maps arguments
 $U_i \cup \{pa_i\}$ to the result
 $V_i$, 
therefore $V_i = h_i(pa_{i,1},pa_{i,2},\ldots,U_i)$; 
the notation  
$pa_{i}=\{pa_{i,1},pa_{i,2},\ldots\} \subset V\setminus V_i$ 
refers to the collection of
endogenous variables entering as argument into $h_i$,
and it is motivated by the qualitative 
representation of these mappings through the directed acyclic graphs
described in this section.
In the following analysis,
 we  assume that the  exogenous variables in $U$ are marginally independent,
thus,  the joint distribution is:
$$p(u_1,u_2,\ldots, u_r ,\ldots,u_{n_V})=\prod_{r=1}^{n_V} p(u_r)$$
where $n_V = \lvert V \rvert$ is the total number of nodes.
Our expert believes this is a reasonable  assumption 
at the considered level of model granularity.
In other terms,  endogenous variables are explicitly and causally
related, while exogenous variables incorporate all residual causal determinants.
NP-SCMs are structural because (implicit) deterministic functions relate
exogenous variables; they are
modular because the  intervention on a variable $V_i$ implies
a local change of  the manipulated variable, i.e., 
its deterministic function $h_i$
becomes equal to the constant assigned by intervention.

\begin{figure}
\begin{center}
\includegraphics[width=4.5in]{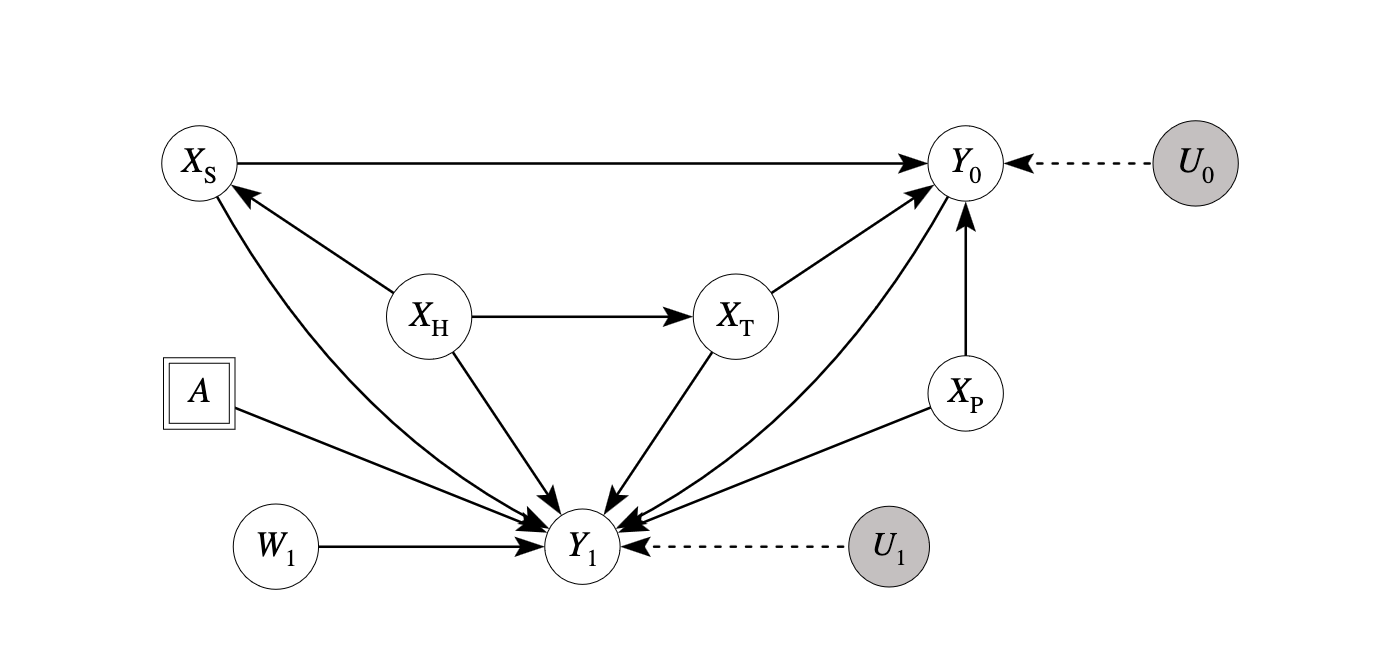}
\end{center}
\caption{DAG $\mathcal{G}_1$ for the minimal observational study on a soldering process
with  regime indicator $A$ and exogenous variables $U_0,U_1$ (grey background)
acting on  $Y_0$ and $Y_1$  made explicit.
\label{fig:DAG_01}}
\end{figure}

A DAG $\mathcal{G}$  is defined by a triple $\mathcal{G}= \langle V,U, E \rangle$,
and it  qualitatively represents the causal relationships
in a SCM   \citep{Pearl2009}.
The nodes in the two sets $V$ and $U$ correspond to endogenous and exogenous variables respectively.
The set $E$ contains directed edges
that join   arguments in the  deterministic function $h_i \in \mathcal{H}$,
called parents 
$pa_i  =  \{pa_{i,1},pa_{i,2},\ldots\}$,  
to the endogenous variable  $V_i$.
An oriented edge, say $X_S \rightarrow Y_1$, represents a
 direct causal relationship  of $X_S$ on $Y_1$
(e.g., Figure~\ref{fig:DAG_01}) at the considered level of
model granularity.
All endogenous variables are represented in a causal DAG, while (some) exogenous
variables are often omitted
unless they are essential at a particular stage of the analysis.

In Figure (\ref{fig:DAG_01}), 
the minimal causal DAG consisting  of only one measurement time $W_1$
is decorated to improve the representation.
Dashed arrows join exogenous to endogenous variables $Y_0$ and $Y_1$, while solid lines always pertain endogenous-to-endogenous causal relationships. 
The grey background of    $U_0$ and $U_1$ 
emphasizes that they are exogenous random variables, while
all  endogenous  variables
have white background to highlight that
realizations   depend on   the value taken
by all parent nodes according to the underlying deterministic functions.
The regime variable $A$  is located within a rectangle
to make explicit that it is a known attribute, i.e.,
it is always known  which regime  is in force.

 In the causal DAG  $\mathcal{G}_1$
(Figure \ref{fig:DAG_01}), 
the electrical  resistance $Y_1$ measured at time $W_1$
depends on the initial electrical resistance $Y_0$ and on covariates
$X_S,X_T,X_P$, besides humidity $X_H$.
The level of   humidity  $X_H$ also determines the
probability that the electronic device 
is configured as $(x_S,x_T)$ by the end user,
e.g., a specific type of component and surface finish
is favored when humidity is high.
Variable $X_H$ satisfies the back-door criterion
\citep[theorem 3.3.2]{Pearl2009}, 
thus in this first simple observational study
the causal effect  of $X_S$ on $Y_1$ conditional on $w_1$
is obtained as:
$$ p(y_{1} \mid do(x_S),w_1)= 
\sum_{x_H} p(y_{1} \mid x_S,w_1,x_H)~ p(x_H)
$$
where the estimand on the left is rewritten
in terms of quantities available from an observational study:
confounding bias is removed.

\begin{figure}
\begin{center}
\includegraphics[width=3.5in]{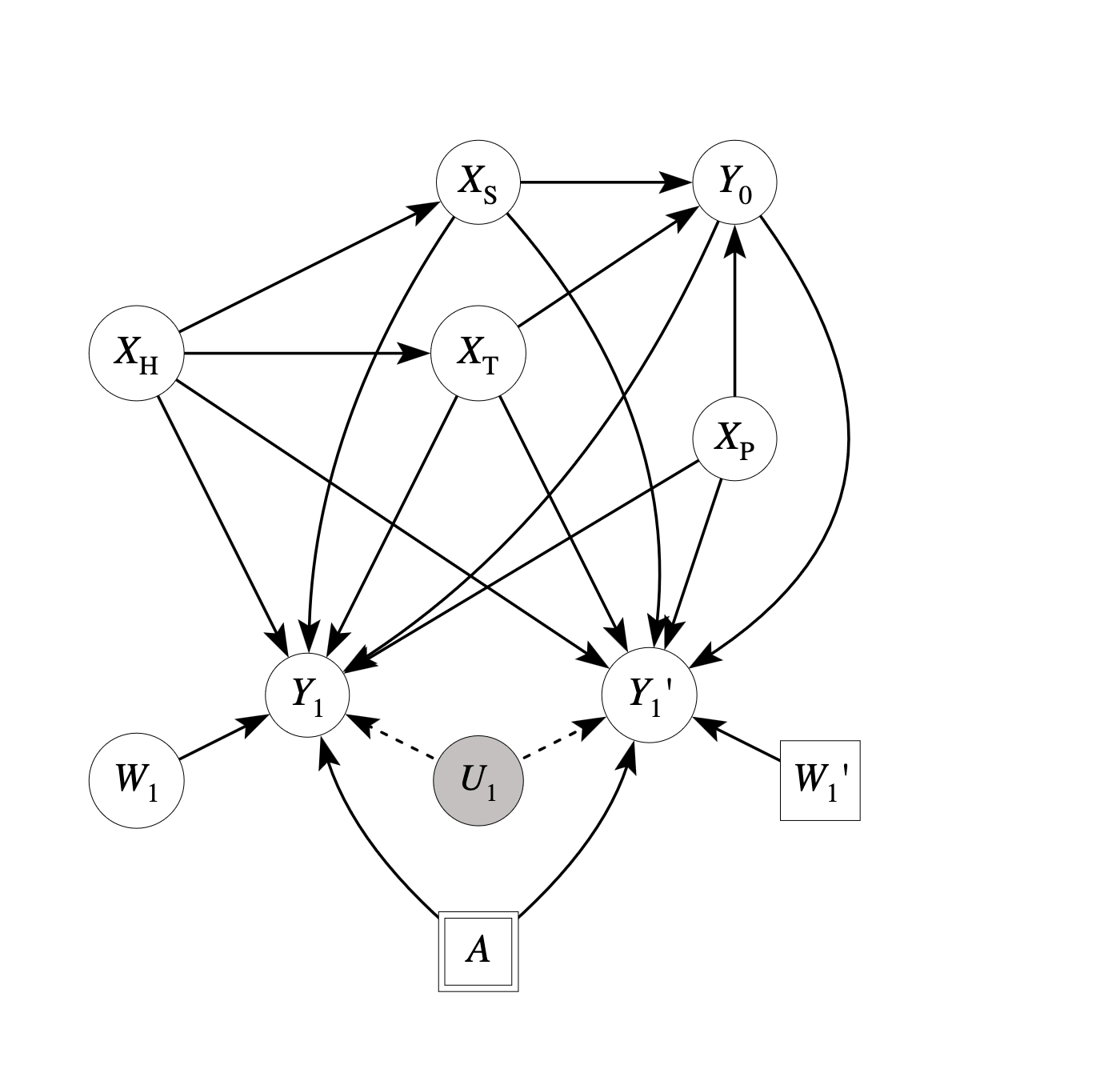}
\end{center}
\caption{DAG $\mathcal{G}_2$ of the twin network (modified) based
on $\mathcal{G}_1$
for what could have happened if  $W_1$ in electronic device $j$ had
been  equal to $w'$ instead of the    observed value $w$.
\label{fig:DAG_02}}
\end{figure}

The counterfactual variable $Y_{1,j,w_1}(w')$ is defined as the value of electrical resistance that would have been measured at time $W_{1,j} = w'$ in the electronic component $j$,
whereas the actual measurement was taken at time $W_{1,j} = w$,   yielding
the observed value $Y_{1,j}=y$.
A common notation for counterfactual variables assigns a subscript to each manipulated variable.
For example,
$Y_{1,j,w_1}(w') = h_{Y,1}(y_{0,j}, x_{S,j},x_{T,j},x_{P,j},x_{H,j}, w', a, u_{1,j})$, 
where $u_{1,j}$ denotes the realization of the exogenous variable $U_{1,j}$
in the factual (i.e., observed) world  in which 
$W_{1,j} = w$ and $y_{1,j}$ are measured.

Twin networks \citep[section 7.1.4]{Pearl2009} are  DAGs representing factual and counterfactual worlds. 
In Figure \ref{fig:DAG_02}, the background nodes shared between the factual and counterfactual worlds 
are not duplicated  \citep{RuizTagle2022}. 
Therefore, only the nodes $Y_1$ and $W_1$,  which differ across worlds, are replicated in the counterfactual representation.
This DAG shows the way to solve the question about the increase of
electrical resistance due to a different
amount of elapsed time in this simple context, but it may  also be  useful
to answer the following question:
which is the amount of  target time $W_{1,j}=w''$ that I should have waited for the electronic device $j$ in order to observe 
a value of electrical resistance equal to $1.1\cdot y_{0,j}$,
given that   $y_{1,j}$ was observed at time $w$?
The answer is available if we can solve the deterministic equation
that defines the failure time 
from the equality  
$Y_{1,j,w_1}(w'') - 1.1 \cdot y_{0,j} = 0$ given
 the endogenous parents in DAG $\mathcal{G}_2$ (Figure \ref{fig:DAG_02}).
We note in passing that
$u_{1,j}$ is calculated in  the factual
world, using the observed  value of $y_{1,j}$;
furthermore,
not only  epistemic (knowledge-based) uncertainty
but also aleatory uncertainty arising from the inherent randomness of a process
cannot be neglected, especially in the case of limited sample sizes.

\section{A parametric Structural Causal Model}
\label{sec:paraSCM}

\begin{figure}
\begin{center}
\includegraphics[width=4.5in]{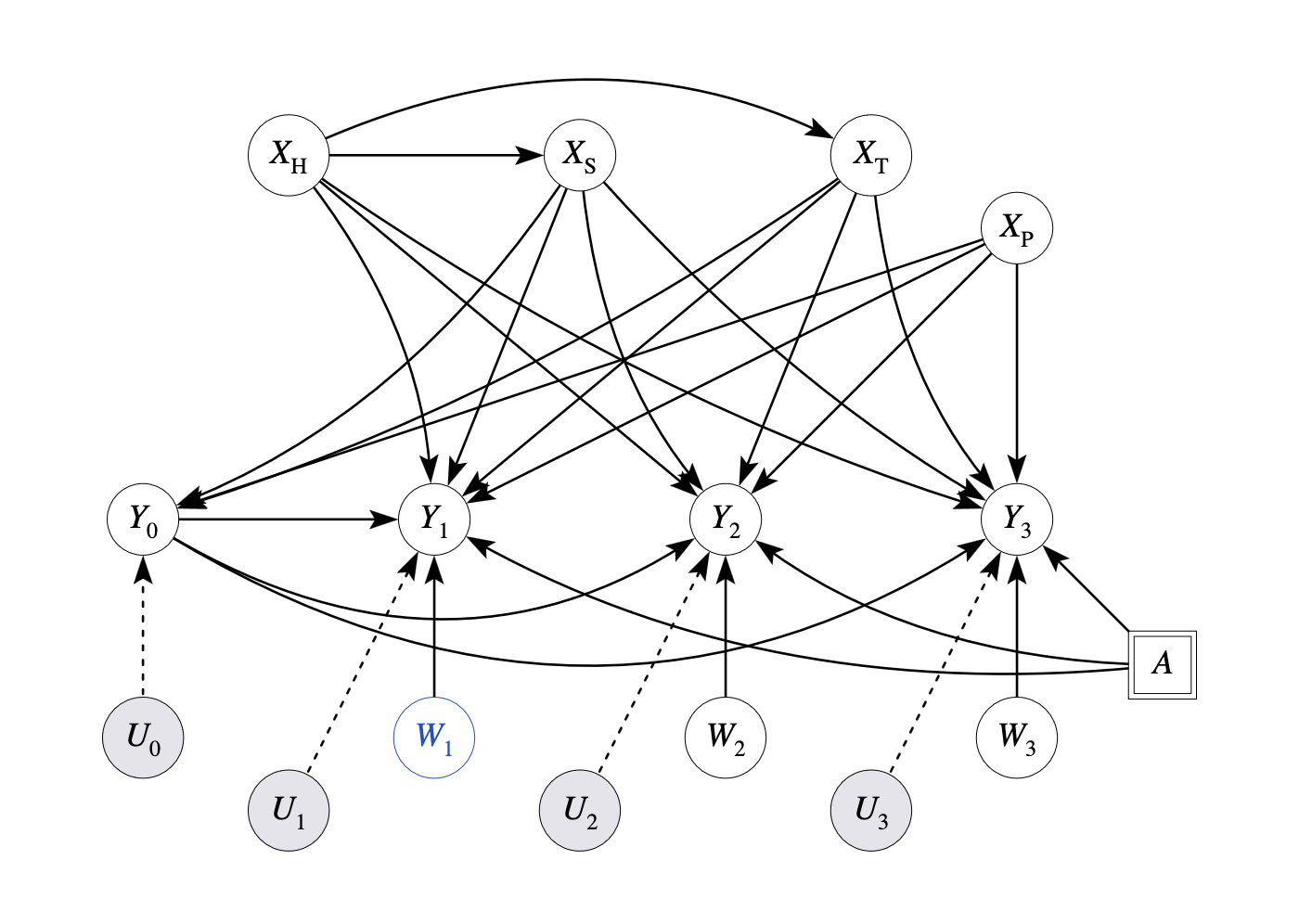}
\end{center}
\caption{DAG $\mathcal{G}_3$ for observational studies
with  regime indicator $A$; exogenous variables are included only when acting on the  electrical resistance $Y_i,~ i=0,1,2,3$ of the electronic device.
 \label{fig:DAGobser2}}
\end{figure}

In this section more details are provided about the considered context 
in which a total of four measurements are  performed on each
electronic  device.
The qualitative representation of the SCM elicited from the expert  is
shown  in  Figure (\ref{fig:DAGobser2}),
and it accommodates both observational studies
under the NS regime and experiments in the AS regime,
where  four measurements are performed including $y_{0,j}$.
The implicit deterministic functions in $\mathcal{H}$
are now refined into parametric models,
where the posterior distribution of unknown parameters given data
is approximated by Markov Chain Monte Carlo (MCMC) simulation.
All exogenous variables are here assumed to be marginally independent.

An observational study, in our context, is   performed by gathering  data 
from a sample of devices installed in production sites worldwide.
Nodes $X_S, X_T, X_P$ are categorical variables representing
the configuration of the current electronic device $j$ in terms of the selected 
features, 
while $X_H = h_H(U_H)$ 
indicates intervals of relative humidity.
Our expert believes that the configuration of each electronic
device is selected by considering the typical humidity 
that   occurs during operation in the assigned production site.
If humidity is higher than $75\%$ for $5$ or more hours in a day
then $X_H =  +1$  indicates  high relative humidity,
otherwise $X_H =  -1$  indicates  normal relative humidity.
After marginalization of the exogenous variable
$U_H$, the induced  binary random variable  $X_H$ is
defined as:
\begin{subequations}\label{eq:XH}
\begin{align}
p(x_H \mid \boldsymbol{\pi}_H) = & 
   \sum_{i\in \{-1,+1\}} \pi_{H,i} ~I_{(i)}(x_H)\\
\Omega_{X_H} =& \{-1,+1\} \\
  \pi_{H,1} =& 1- \pi_{H,-1}  
\end{align}
\end{subequations}
with $I_{(i)}(x)$ the indicating function equal to $1$ if $x = i$,
zero otherwise.
The number of pins is here modeled as a categorical variable
after  integrating out its exogenous variable $U_P$:
\begin{subequations}\label{eq:XP}
\begin{align}
p(x_P \mid \boldsymbol{\pi}_P) = & \sum_{i=1}^{n_P} \pi_{P,i} ~I_{(i)}(x_P)\\
\Omega_{X_P} =& \{1,2,\ldots,n_P\}  \\
\sum_{i=1}^{n_P} \pi_{P,i} = &1
\end{align}
\end{subequations}
where  the notation parallels what defined for $X_H$.

The endogenous variables $X_S=h_S(x_H,U_S)$ and $X_T= h_T(x_H,U_T)$
receive an edge from $X_H$
(Figure \ref{fig:DAGobser2}), and 
the  induced random variables conditional on $X_H$ are
defined as follows:
\begin{subequations}\label{eq:XSXT}
\begin{align}
p(x_S \mid \boldsymbol{\pi}_{S},x_H=i) = &
    \sum_{i=1}^{n_H} I_{(i)}(x_H)   \sum_{r=1}^{n_S} \pi_{S,i,r} ~I_{(r)}(x_S)\\ 
\Omega_{X_S} =& \{1,2,\ldots,n_S\}\\
\sum_{r=1}^{n_S} \pi_{S,i,r} = & 1 ~~~\forall i\in \{1,\ldots,n_H\} \\
p(x_T \mid \boldsymbol{\pi}_{T},x_H=i) = &
    \sum_{i=1}^{n_H} I_{(i)}(x_H)   \sum_{r=1}^{n_T} \pi_{T,i,r} ~I_{(r)}(x_T)\\ 
\Omega_{X_T} =& \{1,2,\ldots,n_T\} \\
\sum_{r=1}^{n_T} \pi_{T,i,r} =& 1 ~~~\forall i\in \{1,\ldots,n_H\}
\end{align}
\end{subequations}
thus  matrices $\boldsymbol{\pi}_{S}$ and $\boldsymbol{\pi}_{T}$
have rows adding to one.

Two remarks are due, although already mentioned
for the simple DAG
of Section (\ref{sec:scmDAGS}).
First, Humidity $X_H$
plays a special role working with observational data
because it is  a parent of nodes $X_S, X_T$
and   also a parent of the outcome $Y_t$, for  all time points $t > 0$. 
In other words, $X_H$ is a confounding variable in the    estimate of
the  causal effects of   $X_S$ and $X_T$   on $Y_t$
at a specified value of time $w_t$.
Second, the node $A$   is represented by a double-rectangle to emphasize its
nature of known regime indicator, i.e., a context
 variable without exogenous input.
Edges $A \rightarrow Y_t$, with $t>0$, make explicit the dependence
of the electrical resistance on the adopted regime.
In all observational studies $A=a_1$ because a NS regime is in force.

The measurement time since a device was turned on
is represented by nodes $W_1, W_2,W_3$ if
no more than three appointments are feasible in production.
In Figure (\ref{fig:DAGobser2}), variables
$W_1, W_2, W_3$  are inside  circular nodes
because the exact time of measurement might be unrecorded,
but for device $j$  it is known that such value is close
to the nominal value $\mu_{W,i}$.
Therefore, the structural specification
$W_{r,j} = h_r(U_{W,r,j})$, for  $r\in \{1,2,3\}$, is refined into
a parametric  specification by  quantifying the  expert's uncertainty as follows:
\begin{align}
\begin{split}\label{eq:uwi}  
 w_{r,j}=&  \mu_{W,r} + ((u_{W,r,j} - 1/2)*1/100), ~ r \in \{1,2,3\}\\
 U_{W,r,j} \sim & Beta(5,5) 
\end{split}
\end{align}
thus a symmetric distribution  is defined whose mode is $\mu_{W,r}$ ranging
around  the nominal value by five hours, i.e.,
$\mu_{W,r} \pm 0.005$.
In our synthetic case study, we consider  $30, 90, 150$ days,
thus $(\mu_{W,1},\mu_{W,2},\mu_{W,3}) = (0.72,2.16,3.60)$
kilohours.

The core component of our Bayesian SCM is 
a parametric model describing electrical resistance across
four observation times.
The equation $Y_0 = h_{Y_0}(x_S,x_P,x_T,U_0)$
takes the following  parametric form:
\begin{align}
\begin{split}\label{eq:y0}
y_{0,j}=&  \mu_0 + 
          \sum_{r=1}^{n_S} \alpha_{S,r}~ I_{(r)}(x_{S,j}) +
          \sum_{r=1}^{n_T} \alpha_{T,r}~ I_{(r)}(x_{P,j}) +
          \sum_{r=1}^{n_P} \alpha_{P,r}~ I_{(r)}(x_{T,j}) +
          u_{0,j}
\end{split}
\end{align}
where $\mu_0$ is the marginal mean and it is expected
to be equal to the nominal value of electrical resistance
for the considered devices (e.g., $1000$ ohms).
Given the sum-to-zero  parameterization,   alphas in equation (\ref{eq:y0})
for  the same endogenous variable add to zero,
e.g., $ \sum_{r=1}^{n_S} \alpha_{S,r} = 0$.
The random variable $U_{0,j} \sim N(0,\sigma^2_0)$
represents  the overall contribution
of all  causal determinants not accounted for as
endogenous variables
to  the initial electrical resistance $Y_0$.
As regards nodes $Y_{t}$ with $t\in\{1,2,3\}$,
the non-parameteric specification
$y_{t} = h_{Y_t}(y_0, x_S,x_P,x_T,x_H, w_t,a',u_t)$
is refined into the following parametric equation:
\begin{subequations}\label{eq:yt}
\begin{align}\label{eq:ytaa}
y_{t,j} = &  y_{0,j} +\\
  \label{eq:yta}
 +& \left(\beta_1 +   
   \sum_{r=1}^{n_S} \delta_{1,S,r}~ I_{(r)}(x_{S,j}) +
   \sum_{r=1}^{n_T} \delta_{1,T,r}~ I_{(r)}(x_{T,j}) +
   \sum_{r=1}^{n_P} \delta_{1,P,r}~ I_{(r)}(x_{P,j}) + \right.\\ \label{eq:ytb}
  &\left. + \sum_{r=1}^{n_H} \delta_{1,H,r}~ I_{(r)}(x_{H,j}) \right)\cdot  
    ~\left[
      w_t \left(I_{\{a_2\}}(a') +\frac{1}{\gamma}I_{\{a_1\}}(a')\right) 
      \right] + \\ \label{eq:ytc}
  & + \left(\beta_2 + 
      \sum_{r=1}^{n_S} \delta_{2,S,r}~ I_{(r)}(x_{S,j}) +
      \sum_{r=1}^{n_T} \delta_{2,T,r}~ I_{(r)}(x_{T,j}) +
      \sum_{r=1}^{n_P} \delta_{2,P,r}~ I_{(r)}(x_{P,j}) + \right.\\ \label{eq:ytd}
  &\left.    
      \sum_{r=1}^{n_H} \delta_{2,H,r}~ I_{(r)}(x_{H,j}) \right) 
   \cdot ~(w_{t,j} -\psi)^{\tau} \cdot I_{\{w_{t,j} -\psi >0\}}(w_{t,j}) \cdot ~I_{\{a_2\}}(a')+\\
  &+ u_{t,j} \label{eq:yte}
\end{align}
\end{subequations}
where deltas referred to the same categorical variable add to zero
(sum-to-zero parameterization),
e.g. $\sum_{r=1}^{n_S} \delta_{2,S,r}=0$;
exogenous random variables are Normal,
$U_{t,j} \sim N(0,\sigma^2_{Y}), ~~t\in\{1,2,3\}$,
with  constant variance, and they represent  the overall
contribution of all  causal determinants not accounted for as
endogenous variables to the electrical resistance $Y_t$. 
Variables   like surface finish and humidity
change the slope of time through parameters
like $\boldsymbol{ \delta_{1,S}} =(\delta_{1,S,1},\ldots,\delta_{1,S,n_S})$
and  $\boldsymbol{\delta_{1,H}}=(\delta_{1,H,1},\ldots,\delta_{1,H,n_H})$
(equation \ref{eq:yt}, lines \ref{eq:yta} and \ref{eq:ytb}),
while  parameters like $\boldsymbol{ \delta_{2,S}} =(\delta_{2,S,1},\ldots,\delta_{2,S,n_S})$
and  $\boldsymbol{\delta_{2,H}}=(\delta_{2,H,1},\ldots,\delta_{2,H,n_H})$
(equation \ref{eq:yt}, lines \ref{eq:ytc} and \ref{eq:ytd})
change the importance of the non-linear component.
Parameter $\psi$ defines the location of a knot such that,
for $W_t$ smaller than $\psi$, this non-linear component of time  is null
(lines \ref{eq:ytc} and \ref{eq:ytd}),
i.e., if  $\psi=2.0$ this component is equal to zero during the first $2000$
hours.
The exponent $\tau$ controls the amount of non-linear  change
in the AS regime,  $A=a_2$,
and $\tau=3.0$  seems a plausible value in our context.

Last, the regime indicator $A=a_1$ makes the non-linear component null
because, without acceleration, the change of electrical resistance is linear in time
for all the lifespan of a device.
Furthermore,  the time transformation function 
defined in line (\ref{eq:ytb}) of equation (\ref{eq:yt})
is introduced  because under the NS regime  at time $w_t$ 
the increase of the electrical resistance is $\gamma$ times smaller than the increase under the AS regime at the same time $w_t$.
In the following sections, we assume that   $\gamma =10$,
therefore $1000$ hours in the NS regime are equivalent
to $100$ hours in the AS regime for the
linear-in-time portion of the increase in electrical resistance.
This part of the elicitation, where the 
two regimes are heavily related, depends on the knowledge of the
underlying physics of failure.
Indeed,  the elicited equation is specific
for the considered context, but we
believe that it may accommodate many other contexts
by selecting plausible intervals
of  parameters in equation (\ref{eq:yt}),
like $\gamma, \psi$ and $\tau$.

\begin{figure}
\begin{center}
\includegraphics[width=4in]{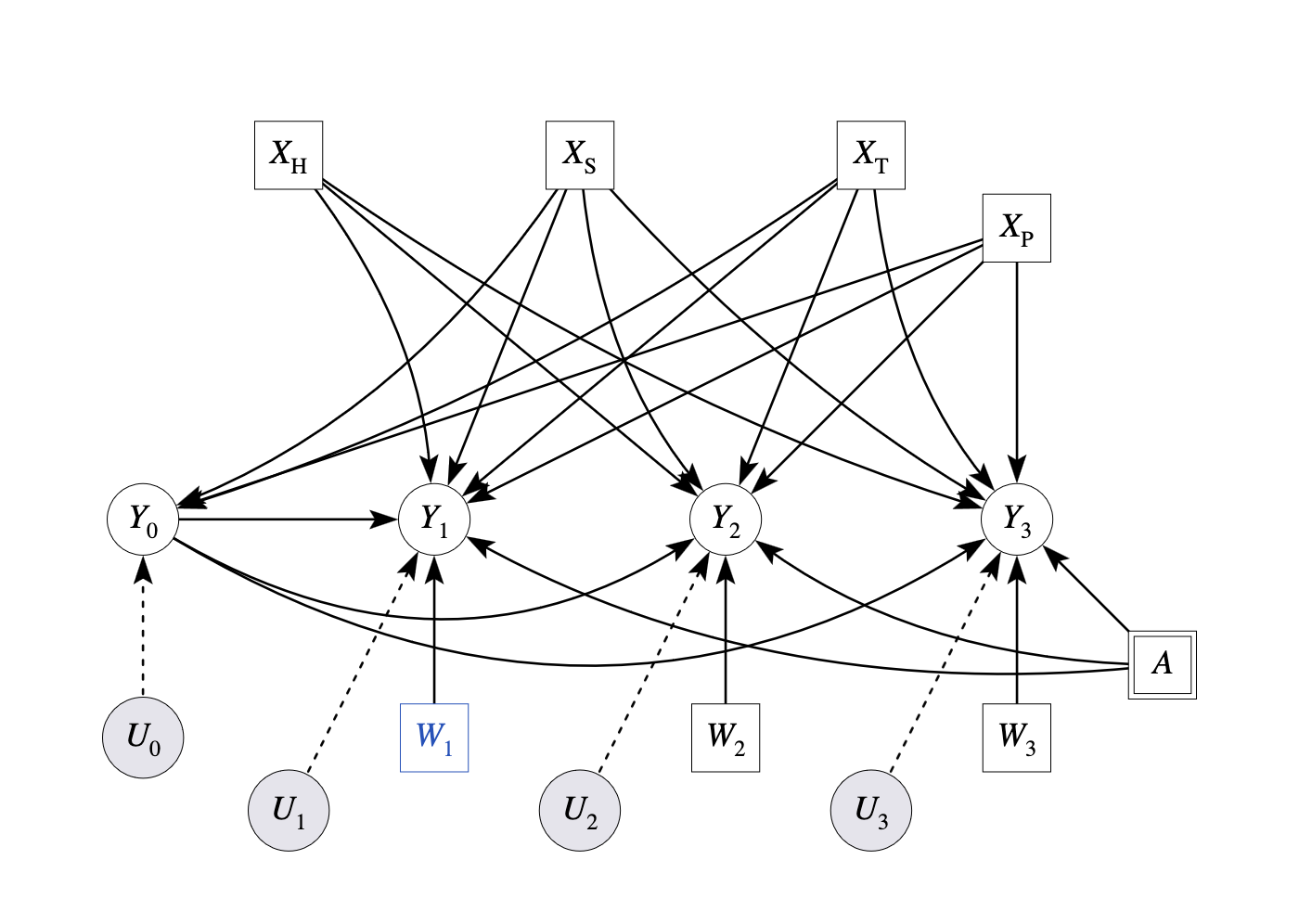}
\end{center}
\caption{Implied DAG  $\mathcal{G}_4$ representing the submodel exploited to estimate
effects of the intervention $do(x_S,x_T,x_P,x_H)$, with $A=a_2$
the  regime indicator.
\label{fig:DAGexpe2}}
\end{figure}

The elicited causal DAG suited  to   observational studies
(Figure~\ref{fig:DAGobser2})  under the NS regime
is also useful for experimental studies in the AS regime after
deriving the implied  submodel (Pearl, 2009, section 7.1.2).
During an experiment, $X_S, X_T, X_P,X_H$ are under full control
of the technologist,
for example humidity is maintained into
a predefined interval  of values and the surface finish is
assigned by randomization.
The edges originated from 
$X_H$ and reaching   nodes $X_S, X_T$ are deleted,
and  $U_H,U_S,U_T,U_P$ are removed
because these nodes do not define values of 
$X_H, X_S, X_T,X_P$ anymore.
If the observation times are  assigned and known without uncertainty  for each device,
then variables $W_t, ~ t>0$, become known constants and all correspondent $U_t$ are
also removed from the DAG.
Nodes $W_1,W_2,W_3$,  
as well as nodes $X_H, X_S, X_T,X_P$, 
are now represented by squares because they are assigned constants. 
After taking   account of the intervention,
the  DAG in  Figure (\ref{fig:DAGobser2})
is transformed into the DAG in Figure (\ref{fig:DAGexpe2}).

\section{Causal estimates leveraging  two synthetic datasets}
\label{sec:causaestisynth} 

A causal estimand is a quantity of interest for the technologist,  a key  question in the considered context.
It may differ according to the stress regime applied to electronic devices besides the specific purpose of the analysis.
 
In this section, two synthetic datasets  $\mathcal{D}_{NS}$ and
$\mathcal{D}_{AS}$ respectively of   size $n_{NS}= 2048$ and $n_{AS}=1024$ at each time point
are considered.
The $\mathcal{D}_{AS}$ dataset is made by $8$ observations-devices 
for each possible configuration of  $X_S,X_T,X_P$ and $X_H$: $8\times 4\times 4 \times 4 \times 2 = 1024$
statistical units.
The $\mathcal{D}_{NS}$ dataset is obtained using   the graphical model as generator of synthetic observations
for a selected configuration of model parameters.
All data are available for download from a dedicated website (see the Online Supplements section).
During the analysis, the expert is certain about   parameters
$\psi=2.00$, $\tau=3$, $\gamma=10$, and the
initial  target electrical resistance 
is set to  $1000$ [ohms].

All statistical analyses were conducted using R \citep{RCoreT2024}
within the RStudio environment \citep{RStudioT2023}. 
Data visualization was carried out using the \textit{ggplot2} package
\citep{Wickham2016},
while Bayesian modeling was implemented
via the \textit{rstan} interface \citep{StanDev2024} 
to the STAN probabilistic programming language \citep{Carpenter2017};
utilities from the \textit{bayetestR} package  \citep{Makowski2019}
were employed to process the MCMC simulation output.

\subsection{Bayesian initial and final distributions}
\label{subsec:bayesIniFin} 

The likelihood function for observational (NS regime)
and experimental (AS regime)
data is defined through the conditional 
probability density functions (pdfs)
defined in  section (\ref{sec:paraSCM}),
then,  a prior distribution is specified for
the model parameters $\boldsymbol{\theta}$.

The posterior distribution of model parameters 
is approximated  by  MCMC
simulation  using the Stan software \citep{StanTeam2025}.

The conditional pdf of endogenous variables given parameters in a PGM is defined 
according to the elicited DAG,
thus the likelihood function of conditionally exchangeable observations is defined as:
\begin{align} \label{eq:jointPGM}
p(v_{1,1},v_{2,1},\ldots,v_{n_G,n} \mid \boldsymbol{\theta}) = 
  \prod_{j=1}^{n}~ \prod_{i=1}^{n_V}  
 ~ p(v_{i,j} \mid pa_i,\boldsymbol{\theta})
\end{align}
where $pa_i=\{pa_{i,1},pa_{i,2},\ldots\}$ is the vector of the parent variables of $v_i$.
Notably, square nodes do not contribute
to equation (\ref{eq:jointPGM}) because they are known constants, like the regime indicator $A$.

In the observational context (NS regime),
the DAG in Figure (\ref{fig:DAGobser2}) is exploited after
changing   $W_1,W_2,W_3$ into square  nodes (known without uncertainty) and modifying equation (\ref{eq:jointPGM})
accordingly.

Prior distributions of  parameters
in the pdfs of  $X_H,X_P,X_S,X_T$ and $Y_0,Y_1,Y_2,Y_3$
are defined during the elicitation step.
Under the assumption that a Dirichlet family of pdfs is suited to represent
the expert's degree of belief about categorical variables,  the
 initial (prior) distributions are defined as:
\begin{align}\label{eq:piSTP}
p(\boldsymbol{\pi}_{S,r} \mid x_H=r,\boldsymbol{\lambda}_{S,r}, a_1) =   Dirichlet(\boldsymbol{\lambda}_{S,r}) \\
p(\boldsymbol{\pi}_{T,r}\mid x_H=r,\boldsymbol{\lambda}_{T,r},a_1) =   Dirichlet(\boldsymbol{\lambda}_{T,r})  \\
p(\boldsymbol{\pi}_{P}\mid \boldsymbol{\lambda}_{P},a_1) =   Dirichlet(\boldsymbol{\lambda}_{P})  \\
p(\boldsymbol{\pi}_{H} \mid \boldsymbol{\lambda}_{H},a_1) =   Dirichlet(\boldsymbol{\lambda}_{H})      
\end{align}
with $r \in \{1,2,\ldots,n_H\}$ 
and where $n_S,n_T$ are respectively the number of
elements in $\boldsymbol{\lambda}_{S,r}$ and 
$\boldsymbol{\lambda}_{T,r}$.
Standard elicitation methods exist to specify the elements of the lambda vectors in the initial distribution \citep{OHagan2006}; however, in this work, all elements are set to one.

The exogenous variable pertaining to a measured electrical resistance $Y_t$
follows a zero-mean Normal distribution with constant variance
for $U_t,~ t>0$.
The prior distribution recommended by \citet{Gelman2006}
for the standard deviation of a Normal distribution
belongs to the half-Student  family, so that the shape is quite flat on the right tail and
with  its maximum at zero, as described in the following:
\begin{align}\label{eq:UT}
U_{t,j} \sim N(0,\sigma^2_{U,t})\\
\sigma_{U,t} \sim \frac{Student(3,0,2.5)}{1-\int_{-\infty}^0 Student(t) ~dt}
\end{align}
with the elicited scale parameter equal to $2.5$ and $3$ degrees of freedom.

The initial distribution for each coefficient in the polynomial model for $Y_t$ is defined
in the location-scale Student family as follows:
\begin{align}\label{eq:betaY}
\delta_{r,q,b} \sim Student(3,0,25)\\
\beta_g \sim  Student(3,0,50)
\end{align}
with $g \in \{1,2\}$,  $r \in\{1,2\}$, $q\in\{S,T,P,H\}$
and $b \in \{1,2,\ldots, n_q\}$, like in equation (\ref{eq:yt}).
Plausible values of these coefficients belong to the interval $(-50,50)$.

The initial distribution of parameters in the linear predictor
of the initial electrical resistance $Y_{0}$ are also in the Student family:
\begin{align}\label{eq:alphaY0}
\alpha_{q,r} \sim Student(3,0,25) \\
\mu_{0} \sim  Student(3,1000,1000)
\end{align}
with $q\in\{S,P,T\}$ and $r\in \{1,\ldots,n_q\}$.

The joint initial distribution of model parameters is defined as:
\begin{align}\label{eq:jointpriorOBS}
p(\boldsymbol{\theta}\mid a_1,\xi)=&
 p(\boldsymbol{\pi}_{P}\mid \boldsymbol{\lambda}_{P})  \cdot
 p(\boldsymbol{\pi}_{H} \mid \boldsymbol{\lambda}_{H})\cdot
 p(\boldsymbol{\alpha}) \cdot p(\boldsymbol{\sigma})
 \cdot p(\boldsymbol{\delta})\cdot p(\boldsymbol{\beta}) \cdot \nonumber\\
\cdot&  \prod_r p(\boldsymbol{\pi}_{S,r}\mid x_H=r, \boldsymbol{\lambda}_{S,r})\cdot 
 \prod_r p(\boldsymbol{\pi}_{T,r}\mid x_H=r, \boldsymbol{\lambda}_{T,r})
\end{align}
where priors for each vector of parameters have been already defined.
Unless otherwise specified, (vectors of) parameters are marginally independent.

In an  experimental study under the AS regime, $A=a_2$, 
randomized treatment assignment defines which 
levels $x'_S, x'_T, x'_P$ of  the experimental factors  are assigned to a (random) sample of devices.
It follows that data in $\mathcal{D}_{AS}$
provide information about variables $Y_0,Y_1,Y_2,Y_3$.
Equation (\ref{eq:jointPGM}) is instantiated
accordingly and the  prior distribution in this context
is limited to:
\begin{align}\label{eq:jointpriorEXPE}
p(\boldsymbol{\theta}\mid a_2, \xi)=&
 p(\boldsymbol{\alpha}) \cdot p(\boldsymbol{\sigma})
 \cdot p(\boldsymbol{\delta})\cdot p(\boldsymbol{\beta}) 
\end{align}
as can be appreciated in DAG (\ref{fig:DAGexpe2}),
after considering the implied likelihood function.

Bayesian model fitting is performed at first by running a MCMC
simulation  for $\mathcal{D}_{AS}$ using the initial
pdf (\ref{eq:jointpriorEXPE}),
and then for $\mathcal{D}_{NS}$ using as prior distribution
equation (\ref{eq:jointpriorOBS}).
Table \ref{tab51} summarizes the results about model parameters in equation
(\ref{eq:yt}) under both the AS and NS regimes (i.e., posterior estimates, posterior standard deviation, and $95\%$ Highest Posterior Density Intervals).
For the sake of brevity, we consider the surface finish coefficients $\delta_{1,S,1}$, $\delta_{1,S,2}$, $\delta_{1,S,3}$, and $\delta_{1,S,4}$. 
Under both the AS and the NS regimes, the corresponding posterior estimates are very close to the true parameter values, with very low posterior standard deviations. 
Satisfactory results are also obtained for the remaining coefficients in equation (\ref{eq:yt}).

\begin{table}[!ht]
\caption{Summary of model (\ref{eq:yt}), surface finish parameters: posterior estimates, posterior standard deviation, $95\%$ Highest Density Interval-HDI.}
\label{tab51}
\begin{tabular}{|c|c|c|c|c|c|}
\hline 
Regime &Symbol & True  value & Post. estimate & Post. Std. Dev.& $95\%$ HDI\\ 
\hline 
AS   & &  &  &  & \\ 
\hline 
&$\delta_{1,S,1}$ & -0.700 & -0.702 & 0.024 & [-0.750, -0.660]   \\ 
\hline 
&$\delta_{1,S,2}$ & -0.500 & -0.493 & 0.024 & [-0.540, -0.450] \\ 
\hline 
&$\delta_{1,S,3}$ & 0.500 & 0.517 & 0.024 & [0.470, 0.570] \\ 
\hline 
&$\delta_{1,S,4}$ & 0.700 & 0.678 & 0.024 & [0.630, 0.720] \\ 
\hline 
NS   & &  &  &  &  \\ 
\hline 
&$\delta_{1,S,1}$ & -0.700 & -0.682 & 0.021 & [-0.720, -0.640]  \\ 
\hline 
&$\delta_{1,S,2}$ & -0.500 & -0.497 & 0.014 & [-0.530, -0.470]  \\ 
\hline 
&$\delta_{1,S,3}$ & 0.500 & 0.510 & 0.016 & [0.480, 0.540] \\ 
\hline 
&$\delta_{1,S,4}$ & 0.700 & 0.669 & 0.016 & [0.640, 0.700] \\ 
\hline 
\end{tabular} 
\end{table}

\begin{figure}[ht!]
\caption{Boxplot of the marginal posterior distributions for
                   parameters $\delta_{1,S,1},\delta_{1,S,2},
                   \delta_{1,S,3},\delta_{1,S,4}$ under the AS regime.}
\label{fig51}
\center
\includegraphics[scale=0.95]{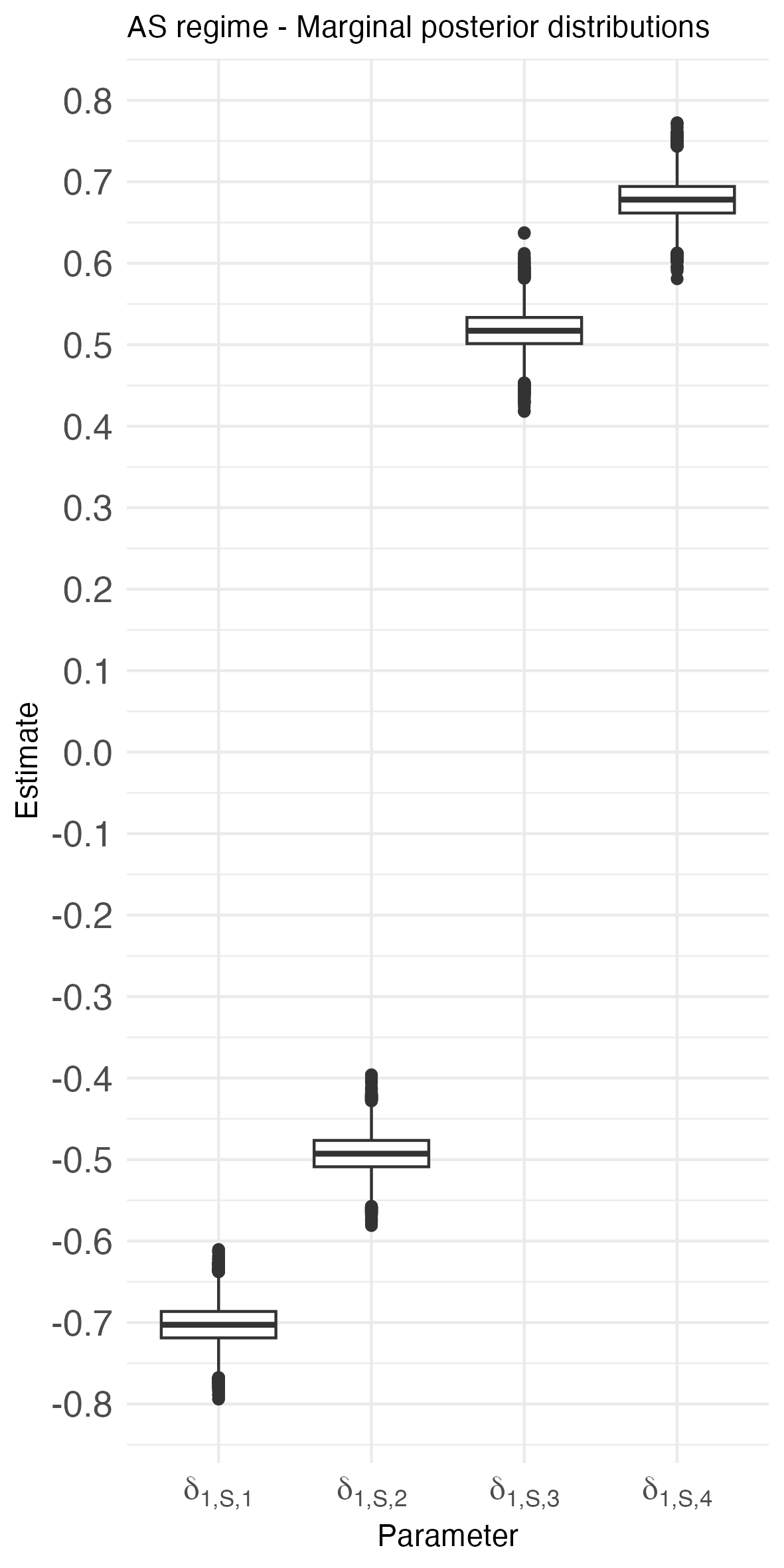} 
\end{figure}

Figure (\ref{fig51})   reports the box-plots of the posterior estimates for the surface-finish parameters under the AS regime. 
The posterior medians are very close to the true parameter values, with very narrow interquartile ranges, confirming the accuracy of the posterior estimates.

\subsection{The increase of electrical resistance in the AS regime}
\label{subsec:increaseAS}

The first estimand quantifies the expected increase
of electrical resistance after $w_t$ hours  of work
for  electronic devices configured as $x_{S}',x_{T}',x_{P}'$
after operating  under  humidity $x_{H}'$:
\begin{subequations}\label{eq:increase}
\begin{align}\label{eq:increaseA}
\Delta_1(x_{S}',x_{T}',x_{P}',x_{H}',w_t,a_2) = 
E[Y_{t} - Y_{0}\mid do(x_{S}',x_{T}',x_{P}',x_{H}',w_t),a_2] =  \\
=E[Y_{t} \mid x_{S}',x_{T}',x_{P}',x_{H}',w_t,a_2] -E[Y_{0}\mid x_{S}',x_{T}',x_{P}',w_t,a_2]= \\ 
  = (\beta_1 + 
   \delta_{1,S,[x_S']} +
   \delta_{1,T,[x_T']}  + 
   \delta_{1,P,[x_P']}+
   \delta_{1,H,[x_H']}) \cdot w_t +~~~~~\nonumber \\
+ (\beta_2 + 
   \delta_{2,S,[x_S']} +
   \delta_{2,T,[x_T']}  + 
   \delta_{2,P,[x_P']}+
   \delta_{2,H,[x_H']}) 
    \cdot ~(w_t -\psi)^{\tau} \cdot I_{\{w_t -\psi>0\}}(w_t)
\end{align}
\end{subequations}
where $\delta_{1,S,[x_S']}$ is  the parameter corresponding to surface finish $X_S = x_{S}'$ and where a corresponding notation holds for all other deltas;
typical times of measurement are  $w_t \in \{0.00, 0.72, 2.16, 3.60\}$
kilohours.

In Table \ref{tab53} we report an example with the values for
the estimand (\ref{eq:increase}), calculated for two
different configurations of $( X_S, X_T, X_P, X_H)$,
i.e., $( 1, 1, 1, 1)$ and $(2, 1, 1, 1)$.
We evaluate the estimand defined in equation (\ref{eq:increase}) under both configurations, across the following values of $w_t$:
$\{0.72, 1.50, 2.00, 2.16, 2.50, 3.00, 3.60\}$.
The posterior standard deviations for all coefficients are low, and the $95\%$
Highest Posterior Density Intervals are narrow, further confirming the accuracy of the parameter estimates in model (\ref{eq:yt}).

\begin{table}[ht!]
\center
\caption{Estimand (\ref{eq:increase})  with two  configurations. Time is expressed in units of $10^{3}$ hours (kilohours).}\label{tab53}
\begin{tabular}{|c|c|c|c|c|}
\hline 
Configuration & Time $w_t$ & Estimate &Std. Dev & $95\%$  HDI  \\ 
$( X_S, X_T, X_P, X_H)$ &  & && \\ 
\hline 
$\lbrace 1, 1, 1, 1 \rbrace$ & 0.72  & 7.330& 0.039& [7.260, 7.410]\\ 
& 1.50 & 15.270 &0.080 & [15.120, 15.430]\\  
& 2.00 & 20.360 &0.107 & [20.150, 20.570]\\  
& 2.16 & 22.116 &0.116 & [21.890, 22.350]\\ 
& 2.50 & 29.324 &0.129 & [29.080, 29.580]\\  
& 3.00 &  61.530& 0.123& [61.290, 61.770]\\ 
& 3.60 & 163.581 &0.116 & [163.350, 163.810]\\  
\hline 
$\lbrace 2, 1, 1, 1 \rbrace$  & 0.72  & 7.481 & 0.038& [7.400, 7.550]\\ 
& 1.50 & 15.585 & 0.080& [15.420, 15.740]\\  
& 2.00  & 20.781 & 0.107& [20.560, 20.990]\\ 
& 2.16 & 22.611 & 0.115& [22.380, 22.830]\\
& 2.50 & 31.101 &0.129 & [30.840, 31.340]\\ 
& 3.00 & 72.170 & 0.123& [71.930, 72.410]\\
& 3.60 & 205.337 & 0.116& [205.110, 205.570]\\
\hline 
\end{tabular} 
\end{table}

\subsection{Increase in electrical resistance under the AS regime with varying configurations}
\label{subsec:deltaincreaseASconf}

The expected increment of electrical resistance $\Delta_1$ may change  according to  alternative configurations of the electronic devices, like levels  $x_{S}' \neq x_{S}''$, and   in general
for different levels of type, $x_{T}' \neq x_{T}''$ and pins,
$x_{P}' \neq x_{P}''$.
The most general contrast of expected increments is a function of
two different tuples of levels
$(x_{S}',x_{T}',x_{P}')$ and
$(x_{S}'',x_{T}'',x_{P}'')$
for the  experimental factors, while
keeping humidity $x_{H}$, time $w_t$ and stress regime $a_2$ at constant values,
as described below:
\begin{subequations}\label{eq:genContra}
\begin{align}\label{eq:genContra1} 
\Delta_{S,T,P}(x_{S}',x_{T}',x_{P}',
               x_{S}'',x_{T}'',x_{P}'',
               x_H,w_t,a_2) =  ~~~~~~~~~~~~~~~~~~~~~~~~~~~~~~~~~~~\\ 
= \Delta_1( x_{S}'',x_{T}'',x_{P}'',x_H,w_t,a_2)
- \Delta_1( x_{S}',x_{T}',x_{P}',x_H,w_t,a_2) =  ~~~~~\\   
= \left[\sum_{q \in \{S,T,P\}}  \sum_{i=1}^{n_q}  \delta_{1,q,i}~ \left(I_{(i)}(x_{q}'') - 
I_{(i)}(x_{q}')\right)\right] ~ w_t ~ +  ~~~~~~~~~~~~~~~~~~~~~~~~~~~~ \nonumber \\
  + \left[\sum_{q \in \{S,T,P\}}  \sum_{i=1}^{n_q}  \delta_{2,q,i}~ \left(I_{(i)}(x_{q}'') - 
I_{(i)}(x_{q}')\right)\right] \cdot
~(w_t -\psi)^{\tau} \cdot I_{\{w_t -\psi>0\}}(w_t)
\end{align}
\end{subequations}
where the sum over index $q$  covers all the experimental factors.

The estimand  in (\ref{eq:genContra1}) includes all instances
of contrast in which just one experimental factor at a time varies.
If only $X_S$ takes two different values  $x_S'$ and $x_S''$,
the estimand becomes:
\begin{multline}\label{eq:increaseS}
\Delta_S(x_{S}',x_{S}'',x_H,w_t,a_2) = \Delta_{S,T,P}(x_{S}',x_{T},x_{P},
               x_{S}'',x_{T},x_{P},
               x_H,w_t,a_2)=\\
= (\delta_{1,S,[x_S'']}-\delta_{1,S,[x_S']}) \cdot w_t +
( \delta_{2,S,[x_S'']}- \delta_{2,S,[x_S']}) \cdot
~(w_t -\psi)^{\tau} \cdot I_{\{w_t -\psi >0\}}(w_t)
\end{multline}
where, $\delta_{2,S,[x_S'']}$ is the   value
to consider for   $X_S=s_{S}''$.
In equation (\ref{eq:increaseS}), $\Delta_S$ is a shortcut notation highlighting
the   experimental factor varied.
Similar causal estimands 
$\Delta_T(x_{T}',x_{T}'',x_H,w_t,a_2)$
and
$\Delta_P(x_{P}',x_{P}'',x_H,w_t,a_2)$
are defined for the other two experimental factors.

Table \ref{tab532} reports the values of estimand (\ref{eq:increaseS}), 
computed using the same set of      $w_t$  values
as   in equation (\ref{eq:increase}). 
Estimand (\ref{eq:increaseS}) represents the one-factor contrast for the two different surface finish values considered when calculating estimand (\ref{eq:increase}) in Table \ref{tab53}. 
The low standard deviations and narrow $95\%$  HPD intervals
confirm that the obtained results are satisfactory.

\begin{table}[ht!]
\center
\caption{Estimand (\ref{eq:genContra}): one-factor contrasts  for two different values of surface finish}\label{tab532}
\begin{tabular}{|c|c|c|c|}
\hline 
 Estimand & Estimate & Std. Dev. & $95\%$  HDI \\ 
\hline 
$\Delta_S(x_{S}',x_{S}'',x_H,w_t,a_2)$ & &&  \\ 
\hline 
$\Delta_S(1,2,1,0.72,a_2)$ & 0.151& 0.028& [0.100, 0.210] \\  
$\Delta_S(1,2,1,1.50,a_2)$ & 0.315& 0.059& [0.200, 0.430] \\  
$\Delta_S(1,2,1,2.00,a_2)$ &  0.420& 0.078 & [0.270, 0.570] \\ 
$\Delta_S(1,2,1,2.16,a_2)$ &  0.495& 0.084 & [0.330, 0.660] \\ 
$\Delta_S(1,2,1,2.50,a_2)$ & 1.777 &  0.094& [1.590, 1.960] \\ 
$\Delta_S(1,2,1,3.00,a_2)$ & 10.640 & 0.086 & [10.470, 10.810]\\ 
$\Delta_S(1,2,1,3.60,a_2)$ & 41.756 & 0.090 & [41.580, 41.940]\\ 
\hline 
\end{tabular} 
\end{table}

\subsection{Reliability of a future device in the AS regime}
\label{subsec:deltaincreaseAS}

A causal effect on reliability (failure time) is defined by the change in
reliability (survival) times under different treatment conditions,
for example a difference in median reliability (survival) time between treated and untreated groups.
The failure time $T$  depends on the difference 
of initial, $Y_{0,j}$  and final, $Y_{t,j}$, values of electrical resistance,
but also on the threshold described as $10\%$ of $Y_{0,j}$, a quantity  
specific for observation (device) $j$.

We now consider a future device that could be added to increase the sample size to $n+1$,
and we maintain   index $j$ to refer to this new statistical unit.
At the beginning of the study, just after the device  is measured,
the random variable to consider is as follows:
\begin{align} \label{eq:regime2diffe0}
D_{0,j} = Y_{t,j} - 1.1 \cdot y_{0,j}
\end{align}
which is a member of the Normal family of pdfs, thus from
equation (\ref{eq:yt}) we have:
\begin{align} \label{eq:pdfregime2diffe0}
p(d_{0,j} \mid y_{0,j},do(x_{S},x_{T},x_{P},x_{H},w_t),a_2) =
            N(d_{0,j}\mid \mu_{D,0,j},~~ \sigma^2_{Y})      
\end{align}
where parameter $\sigma_Y^2$ is the variance of
the exogenous variable $U_{t,j}$, while
$\mu_{D,0,j}$  is equal to the linear predictor r.h.s.
of equation (\ref{eq:yt}), lines (\ref{eq:yta}-\ref{eq:ytd}), after   simplification: 
\begin{subequations}\label{eq:muDOJ}
\begin{align}\label{eq:muDOJa}
\mu_{D,0,j}(w_{t,j})& = h_D(\boldsymbol{x},w_{t,j},\boldsymbol{\theta})-0.1 \cdot y_{0,j} =\\
 =& \left(\beta_1 +   
   \sum_{r=1}^{n_S} \delta_{1,S,r}~ I_{(r)}(x_{S,j}) +
   \sum_{r=1}^{n_T} \delta_{1,T,r}~ I_{(r)}(x_{T,j}) +
   \sum_{r=1}^{n_P} \delta_{1,P,r}~ I_{(r)}(x_{P,j}) + \right.\\ \label{eq:muDOJb}
  &\left. + \sum_{r=1}^{n_H} \delta_{1,H,r}~ I_{(r)}(x_{H,j}) \right)\cdot  
    ~\left[
      w_t \left(I_{\{a_2\}}(A) +\frac{1}{\gamma}I_{\{a_1\}}(A)\right) 
      \right] + \\ \label{eq:muDOJc}
  & + \left(\beta_2 + 
      \sum_{r=1}^{n_S} \delta_{2,S,r}~ I_{(r)}(x_{S,j}) +
      \sum_{r=1}^{n_T} \delta_{2,T,r}~ I_{(r)}(x_{T,j}) +
      \sum_{r=1}^{n_P} \delta_{2,P,r}~ I_{(r)}(x_{P,j}) + \right.\\ \label{eq:muDOJd}
  &\left.    
      \sum_{r=1}^{n_H} \delta_{2,H,r}~ I_{(r)}(x_{H,j}) \right) 
   \cdot ~(w_{t,j} -\psi)^{\tau} \cdot I_{\{w_{t,j} -\psi >0\}}(w_{t,j}) \cdot ~I_{\{a_2\}}(A)+\\
  & -  0.1 \cdot y_{0,j}. \label{eq:muDOJe}
\end{align}
\end{subequations}
with the notation $\mu_{D,0,j}(w_{t,j})$ emphasizing  the dependence
of the expected value on   time $w_{t,j}$.
The reliability function for device $j$ introduced
in equation (\ref{eq:reliability01}) is rewritten in terms of change of electrical
resistance, conditional of the observed time $t_j$, as follows:
\begin{subequations}\label{eq:reliability02}
\begin{align} \label{eq:reliability02a}
R(t_j\mid y_{0,t},a_2,do(\boldsymbol{x}),\boldsymbol{\theta}) = &  P[D_{0,j} < 0 \mid 
W_{t,j}=t_j, y_{0,j},a_2,do(\boldsymbol{x}),\boldsymbol{\theta}] =\\
&=\label{eq:reliability02b}
\int_{-\infty}^{0} N(
q \mid \mu_{D,0,j}(t_j),\sigma^2_{Y}) ~dq
\end{align}
\end{subequations}
with $do(\boldsymbol{x}) = do(x_{S},x_{T},x_{P},x_{H})$;
to avoid ambiguity, we use $q$ as the variable of integration instead of $d_{0,j}$.
If the vector $\boldsymbol{\theta}$  of model parameters is known in
equation (\ref{eq:reliability02a}),  as well as the considered factors,
then the reliability is just a function of
 time $t_j$ (equation  \ref{eq:reliability02b}).

The above analysis is now extended to a future device,
again indexed as $j$, which belongs to the group of devices
with configuration $\boldsymbol{x}$, the only known information about it.
The difference $D_{j} =   Y_j - 1.1 \cdot Y_{0,j}$
conditional on time $w_{t}$, after simplification,
is a random variable with expected value $\mu_D(w_t)$:
\begin{subequations}
\begin{align}
\mu_D(w_t) = & E[h_D(\boldsymbol{x},w_{t},a_2,\boldsymbol{\theta}) + U_{t,j}
- 0.1 \cdot Y_{0,j}] =\\
=&  h_D(\boldsymbol{x},w_{t},a_2,\boldsymbol{\theta}) + E[ U_{t,j}]
- 0.1 ~ h_{0}(\boldsymbol{x},w_{t},\boldsymbol{\theta})
- 0.1 ~E[U_{0,j}] = \\
=& h_D(\boldsymbol{x},w_{t},a_2,\boldsymbol{\theta}) -
   0.1~h_0(\boldsymbol{x},w_{t},\boldsymbol{\theta})
\end{align}
\end{subequations}
since $h_D(\boldsymbol{x},w_{t},a_2,\boldsymbol{\theta})$
does not vary, as well as the linear predictor
$h_0(\boldsymbol{x},w_{t},\boldsymbol{\theta})$,
and  $E[U_{0,j}] = E[U_{t,j}]  = 0$.
The variance of this difference becomes:
\begin{subequations}
\begin{align}
\sigma^2_D = Var[D_j] =& Var[ Y_{t,j} -1.1 ~Y_{0,j}] =\\
=& Var[ h_D(\boldsymbol{x},w_{t},a_2,\boldsymbol{\theta}) +U_{t,j} 
   -0.1 ~h_0(\boldsymbol{x},w_{t},\boldsymbol{\theta}) -0.1~ U_{0,j}] \\
  =&  Var[ U_{t,j}  -0.1 ~U_{0,j}] = \sigma^2_{Y} +0.01 ~\sigma^2_{0}
\end{align}
\end{subequations}
since the variance of constants is null and   exogenous variables
are not correlated; 
given the normality of exogenous variables, the  distribution
of this difference is Normal, $D_j \sim N(\mu_D(w_t),~ \sigma^2_D)$.
We conclude that in this context,  the reliability function is:
\begin{subequations}\label{eq:reliability03}
\begin{align} \label{eq:reliability03a}
R(t_j\mid a_2,do(\boldsymbol{x}),\boldsymbol{\theta}) = &  P[D_{j} < 0 \mid 
W_{t,j}=t_j,a_2,do(\boldsymbol{x}),\boldsymbol{\theta}] =\\
&=\label{eq:reliability03b}
\int_{-\infty}^{0} N(
q \mid \mu_{D}(t_j),\sigma^2_{D}) ~dq
\end{align}
\end{subequations}
with  $\mu_{D}(t_j)$ and $\sigma^2_D$ defined above;
for notational convenience, $q$ is used in place of $d_j$ within the integral.

\begin{figure}[ht!]
    \centering
    \includegraphics[width=0.56\linewidth]{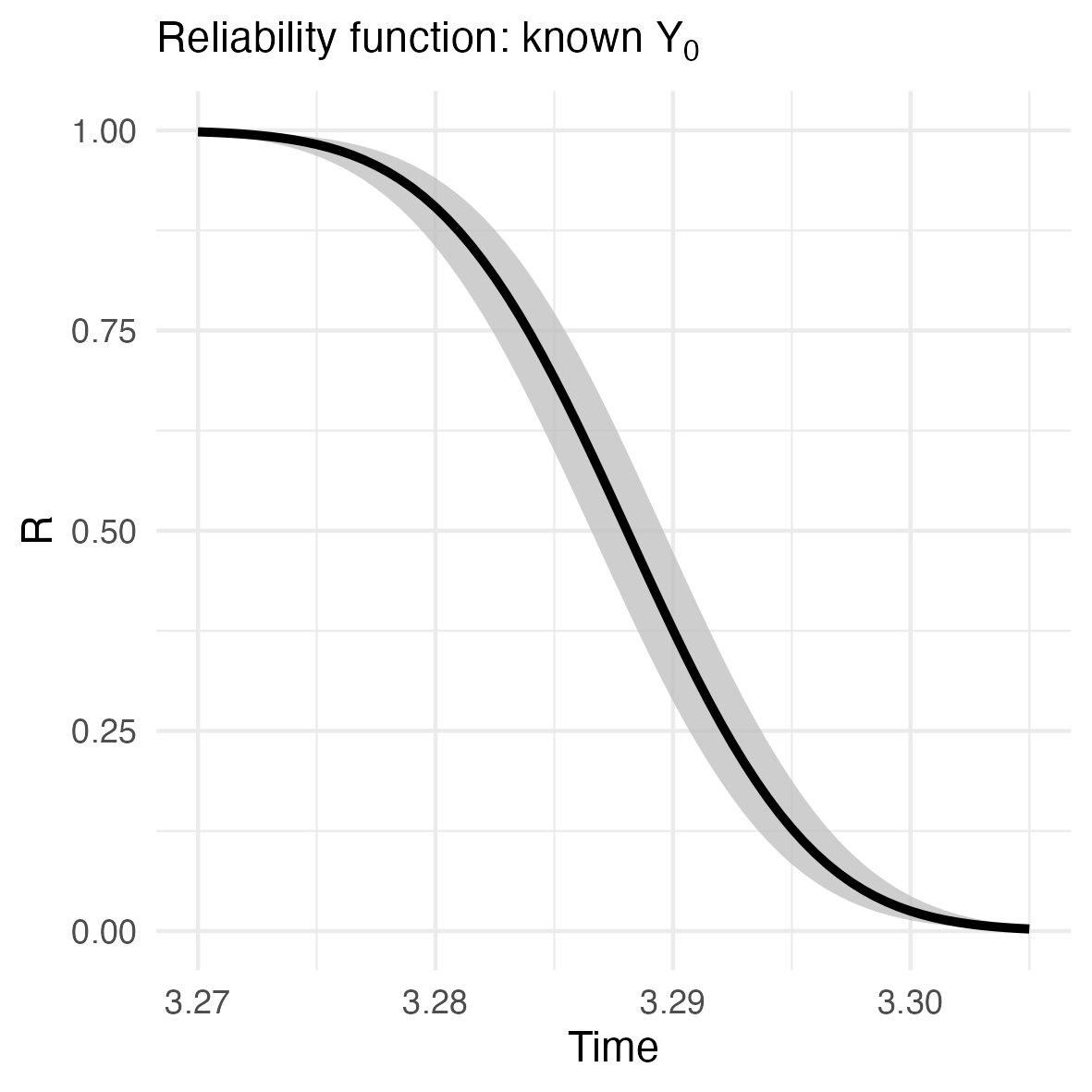} \vspace{0.1cm}
    \includegraphics[width=0.56\linewidth]{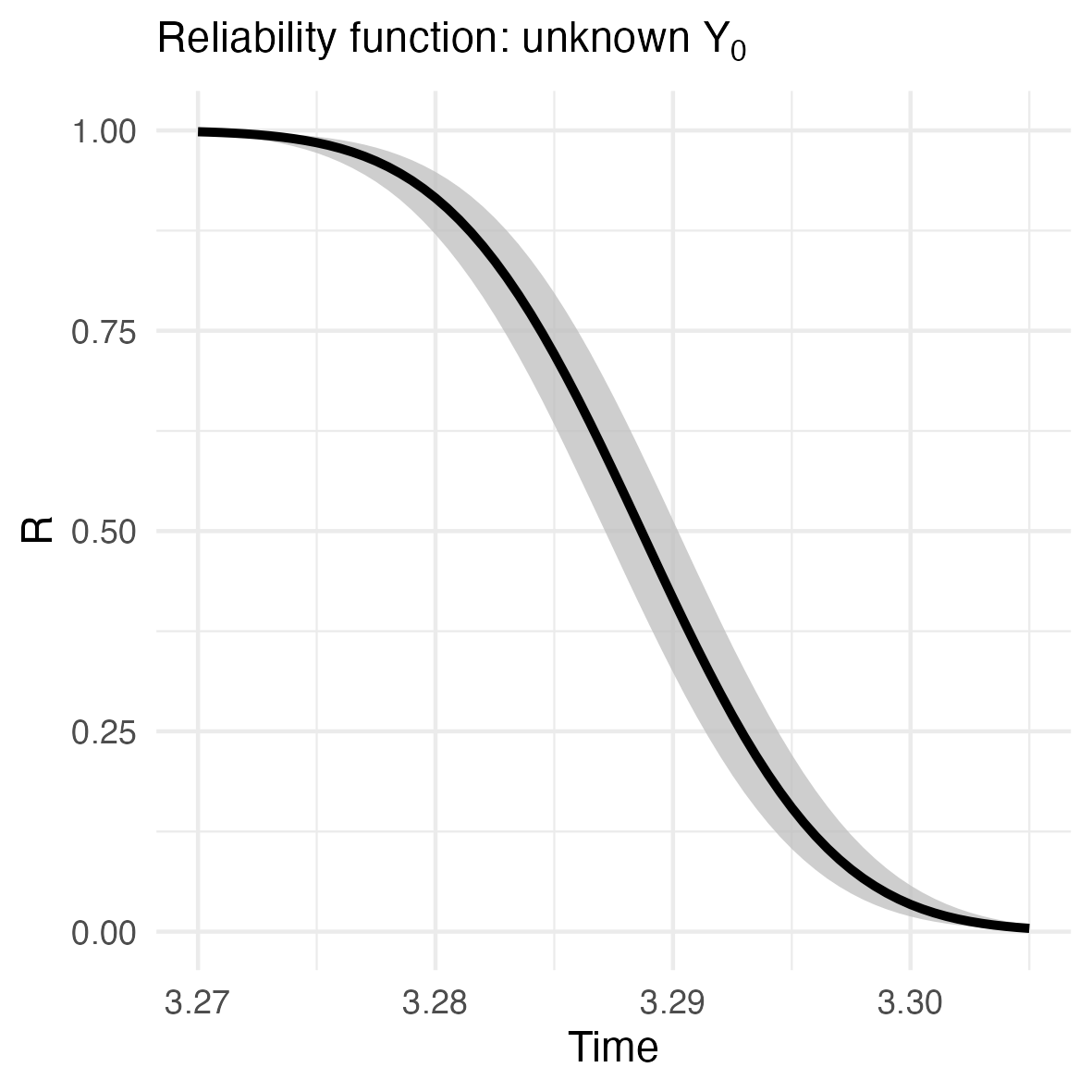}
    \caption{Estimated reliability functions with unknown $Y_0$ (bottom),  equation (\ref{eq:reliability02}), and known $Y_0$ (top),  equation 
    (\ref{eq:reliability03}).  }
    \label{fig54}
\end{figure}

Figure (\ref{fig54}) 
shows the estimated reliability
functions related to formulas 
(\ref{eq:reliability02}) and (\ref{eq:reliability03}), 
with  unknown  (bottom) and   known (top) $Y_0$, respectively. 
Both functions are calculated under the configuration
$( X_S, X_T, X_P, X_H)=( 1, 1, 1, 1)$.

\subsection{Observational studies   in the NS regime}
\label{subsec:causaestimOBS}

\begin{figure}
    \centering
    \includegraphics[width=0.45\linewidth]{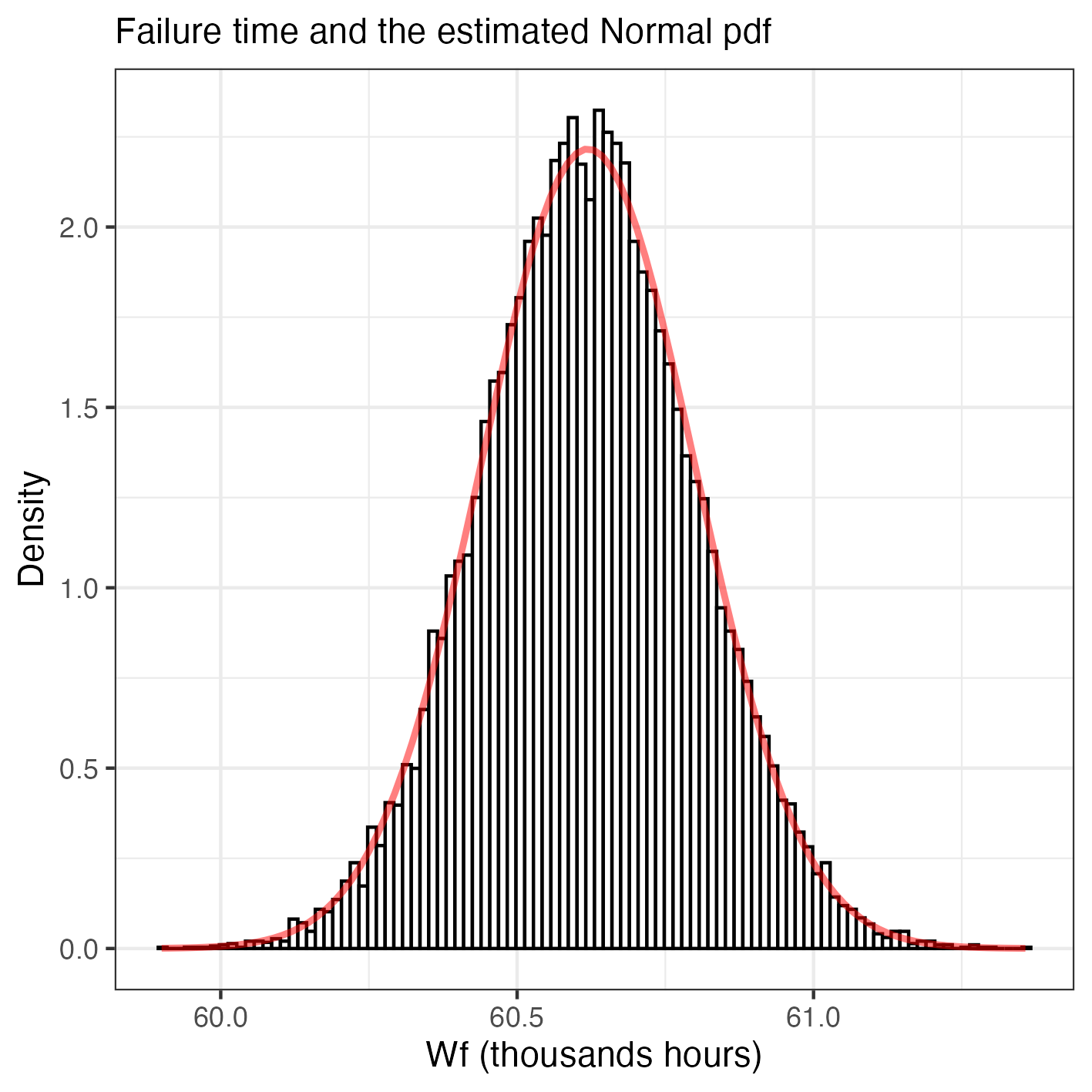}
    \includegraphics[width=0.45\linewidth]{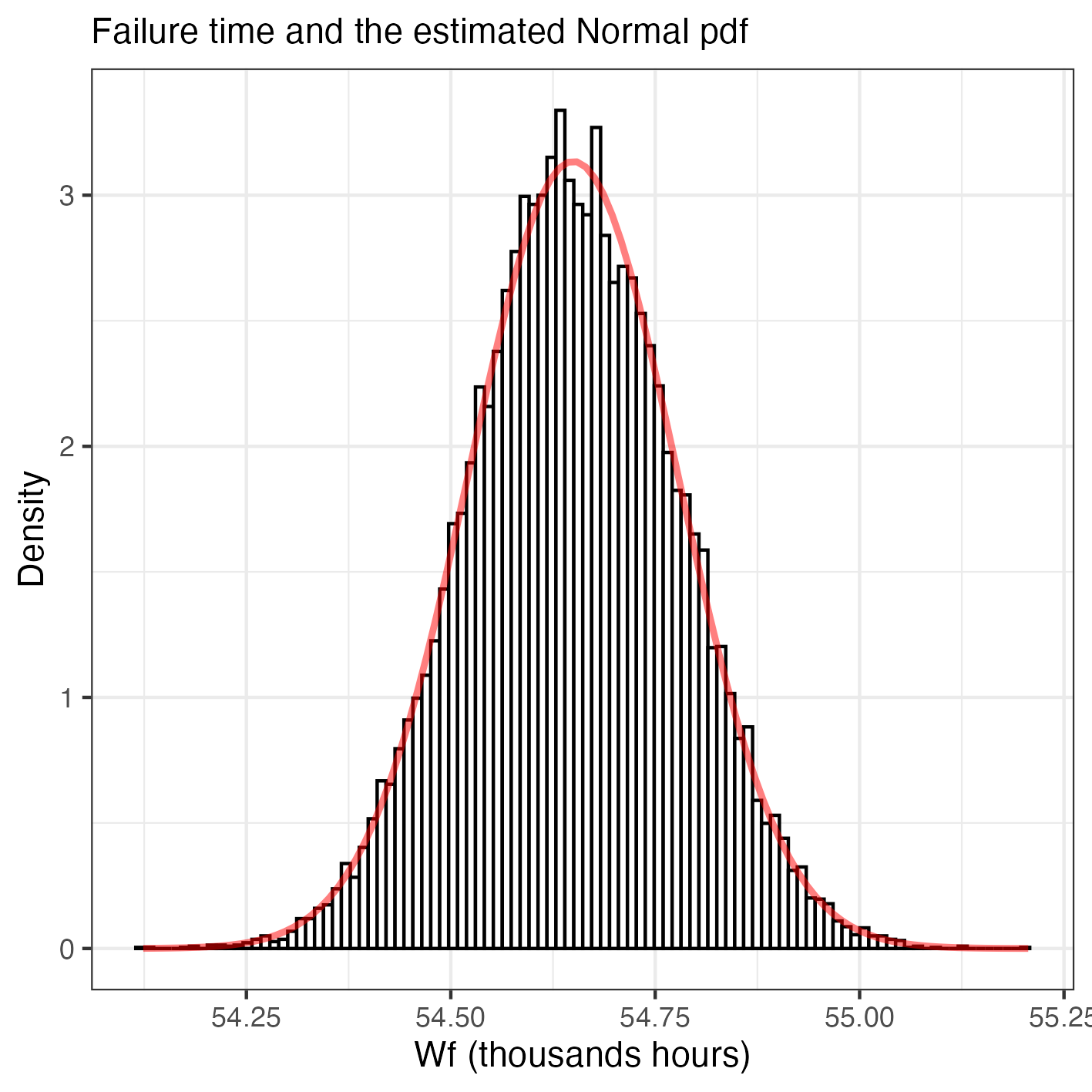}
    \caption{Estimated pdf of $W_f$ in kilohours, given  the intervention $do(x_S=1,x_T=1,x_P=4)$, left,
    and under $do(x_S=3,x_T=3,x_P=3)$, right.}
    \label{fig:NS114}
\end{figure}

Estimates of the causal effects of $X_S$ and $X_T$ on $Y_t$
are biased when based on observational data in the NS regime ($A=a_1$), 
due to the presence of the confounder $X_H$.
Conditioning on $X_H$ suffices to eliminate confounding bias,
as stated by  Pearl's back-door criterion  \citep[sec. 3.3.1]{Pearl2009}.
The truncated product formula is  defined here by removing
all conditional pdfs associated with  the selected intervention
and by marginalizing over the confounder
\citep[sec. 3.2.3 and note 4 page 72]{Pearl2009}.
The resulting expression is as follows:
\begin{multline}\label{eq:regime1joint}
p(y_t,y_0 \mid do(x_{S},x_{T},x_{P}),w_t,a_1) =\\
= p(y_0 \mid x_{S},x_{T},x_{P}) ~
\sum_{x_H}   p(y_t \mid x_{S},x_{T},x_{P},x_{H},y_0,w_t,a_1)~p(x_{H})
\end{multline}
where $do(x_S,x_T,x_P)$ represents the   intervention of interest.
Nevertheless, the failure time $W_f$ of a device is a function of 
$Y_t, Y_0,w_t,\boldsymbol{\theta}$,
thus  the distribution of
the failure time $W_f$ is derived  given a selected intervention,
assuming all model parameters are known, i.e., conditional on $\boldsymbol{\theta}$.
The Bayesian predictive distribution of failure time $W_f$ accounts for the uncertainty about model parameters and is conditional  on
a given surface, type and pin.
A MCMC sample of draws from the predictive distribution
of failure time $W_f$ is exploited to estimate the main features of this distribution.
To illustrate the approach, we consider the intervention
$do(x_S=1,x_T=1,x_P=4)$:
for each parameter value $\boldsymbol{\theta}^{(i)}$ sampled from the posterior distribution at iteration $(i)$, we draw a
realization from the predictive distribution of $p(x_H\mid \boldsymbol{\theta})$, then we draw a realization from
the corresponding component of the mixture distribution.
The same algorithm is followed for the intervention
$do(x_S=3,x_T=3,x_P=3)$.
After obtaining the two histograms (Figure \ref{fig:NS114}),
a Normal pdf is fitted for each intervention,
and displayed as an overlapped continuous line.
Small departures from normality are present in both histograms.
In  Table (\ref{tab55}), summaries of $W_f$ for the 
two considered interventions are shown,
and is clear the huge differences of the two failure time distributions
in the considered population.
The credible interval of level $0.90$ for the first intervention
is $(60.322, 60.913)$ while for the second one is $(54.440, 54.860)$.
Therefore, the two intervals do not overlap: 
the left endpoint of the first interval differs by more than 5,000 hours from the right endpoint of the second one.
Causal estimands   based on contrasts of mean values or ratio of variances
under different interventions may be derived and evaluated
using the  MCMC simulation output.

\begin{table}
\centering
\caption{Summary statistics (kilohours) about the estimated distribution
of the failure time under the NS regime with intervention
(a): $do(x_{S}=1,x_{T}=1,x_{P}=4)$, and (b): $do(x_{S}=3,x_{T}=3,x_{P}=3)$.
\vspace{0.5cm}}\label{tab55}
\begin{tabular}{r|c|c}\hline
Statistic & (a): $do(x_{S}=1,x_{T}=1,x_{P}=4)$       & (b): $do(x_{S}=3,x_{T}=3,x_{P}=3)$   \\ \hline
Min.         &  59.938 &  54.124\\
$w_{f,0.05}$ &  60.322 &  54.440\\
$w_{f,0.25}$ &  60.496 &  54.564\\
$w_{f,0.50}$ &  60.618 &  54.648\\
$w_{f,0.75}$ &  60.741 &  54.736\\
$w_{f,0.95}$ &  60.913 &  54.860\\
Max.         &  61.398 &  55.206\\
Mean         &  60.618 &  54.650\\
Std. Dev.    &   0.180 &  0.127\\ \hline
\end{tabular}
\end{table}

Interval-censored and right-censored  observations of failure time characterize
our context in the NS regime.
If we consider an observation $j$ for which the electrical resistance
is $y_{3,j} > 1.1~ y_{0,j}$ at time $w_{3,j}$, then
we can state that $w_f \in (w_{2,j}, w_{3,j})$, an interval that may span
thousands of  hours.
If the measured   electrical  resistance
$y_{3,j}$ at the latest   time point  $w_{3,j}$
is less than $1.1~ y_{0,j}$, then
we can conclude that $w_f > w_{3,j}$.

A natural question is: at what  time   the equality 
$y_{3,j} = 1.1 ~y_{0,j}$   holds for device $j$?
We rephrase this question in counterfactual terms:
given that the electrical resistance is measured at time 
$w_{3,j}$ and the corresponding observed value is 
$y_{3,j} \neq 1.1 y_{0,j}$,
what time could have been selected instead
to obtain $y_{3,j} = 1.1 ~y_{0,j}$ under the same
assembly configuration and humidity conditions?

In Pearl's  twin-network approach \citep[section 7.1.4]{Pearl2009}, it is possible to obtain
counterfactual quantities  by setting the value of some selected variables 
(Figure  \ref{fig:DAG_02}, section \ref{sec:scmDAGS})
to whatever value is of interest.
Then,  the failure time $W_{f,j}$ for device $j$
may be inferred as a counterfactual quantity:
it is the time $t$   at which the electrical   resistance $y_{t,j}$ of
an electronic  device $j$ would have reached   $1.1 y_{0,j}$ had the observation time $w_t$ been
equal to the solution   of the equation:
\begin{multline}\label{eq:controFa1} 
0.1 \cdot y_{0,j} -\left(\beta_1 +   
   \sum_{r=1}^{n_S} \delta_{1,S,r}~ I_{(r)}(x_{S,j}) +
   \sum_{r=1}^{n_T} \delta_{1,T,r}~ I_{(r)}(x_{P,j}) +
   \sum_{r=1}^{n_P} \delta_{1,P,r}~ I_{(r)}(x_{T,j}) + \right.\\
  \left. + \sum_{r=1}^{n_H} \delta_{1,H,r}~ I_{(r)}(x_{H,j}) \right)\cdot  
   \frac{w_t}{\gamma}  - u_{3,j} = 0 
\end{multline}
in the observational regime, $A=a_1$,
thus:
\begin{equation}\label{eq:teffeR1}
w_{f,j} \mid  \boldsymbol{\theta} = \frac{
0.1 \cdot y_{0,j} -  u_{3,j}
}{ \beta_1+
\delta_{1,S,[j]}+\delta_{1,T,[j]}+
\delta_{1,P,[j]}+\delta_{1,H,[j]}
} \cdot \gamma
\end{equation} 
which is a quantity whose distribution,
conditional on the factual observations
$y_{0,j}, y_{t,j}$,
is governed by the   posterior distribution 
$p(\boldsymbol{\theta} \mid \mathcal{D})$ of model parameters.
It is worth noting that   equation (\ref{eq:teffeR1})
does not contain $y_{t,j}$; nevertheless, 
this value is needed to compute the realized residual
$u_{t,j} \mid \boldsymbol{\theta}^{(g)}$
using  equation (\ref{eq:yt}),
where $\boldsymbol{\theta}^{(g)}$  denotes
the value of the parameter vector   sampled at iteration $g=1,2,\ldots,G$
in the MCMC simulation.
Similarly, the coefficients   in the denominator
are elements in   $\boldsymbol{\theta}^{(g)}$. 
An MCMC sample to approximate   the final distribution
of the counterfactual variable $W_{f,j}$ 
is obtained by iterating the following steps:
i) compute the residual
 $u_{3,j}$ given the current  value  $\boldsymbol{\theta}^{(g)}$ and 
 the observed values $y_{t,j}, y_{0,j}$; 
 ii) calculate $w_{f,j}$
in equation (\ref{eq:teffeR1})  by replacing 
model parameters and the implied residual;
iii) iterate the above steps for $G$ times, the sample size of the MCMC simulation
after warmup.

To illustrate the approach, we consider
  device ID2436, which showed
$Y_{0,j}=1003$ and $Y_{3,j}=1070$ ohms at $w_{3,j}=36$ kilohours.
The distribution of the counterfactual failure time
$W_{f,j}$, approximated by MCMC,
has mean and median of  $53.690$ kilohours, standard
deviation of $0.0185$  kilohours and 5th and 95th percentiles
given by
$w_{f,j,0.05}= 53.660$ and  $w_{f,j,0.95}= 53.720$, respectively.
In other words, in a context where hourly measurements are performed,
the failure time could have been observed at
$\widehat{E}[W_{f,j}] = 53690$ hours, i.e.,
approximately six years of operation.

Counterfactuals are also useful in an hypothetical context of a lawsuit filed
by a client against  the manufacturer of device $j$.
The contract signed by the parties stipulates that at $21.6$ kilohours,
the electrical resistance of device $j$ is almost certainly less than or equal to $1040$ ohms.
However, the client claims to have measured a value of
$1041.056$ ohms.
The manufacturer defends itself by pointing out that the contract also specified
a normal humidity level during operations, whereas during company inspections at the client’s site,
the humidity was consistently found to be high.
What would the measured electrical resistance have been at time $21.6$ kilohours, had the humidity level been kept normal, whereas it was actually kept high and we measured $Y_{2,j} = 1041.056$ ohms?

The answer follows the  aforementioned algorithm, except that step (ii) is replaced by the calculation of the 
counterfactual value $Y_{2,j,x_H}(-1)$,
conditional on $\boldsymbol{\theta}^{(g)}$.
This value is defined as the sum of $y_{0,j}$,
the estimated residual $u^{(g)}_{2,j}$, and the linear predictor including all previously estimated coefficients, except for the one associated with humidity, since the factual value $x_H = 1$ is replaced by the counterfactual value $x_H = -1$.
The sample of $G=20,000$ counterfactual values for
$Y_{2,j,x_H}(-1)$ are all less than or equal to $1034.869$ ohms,
and the credible interval of size $0.98$ has endpoints
$(1034.507, 1034.759)$.
What claimed by the manufacturer is well supported
by the estimated distribution of $Y_{2,j,x_H}(-1)$. 
We note in passing that in a legal context, 'almost sure'
might correspond to a probability value smaller than $1\times 10^{-5}$.

\section{Final remarks and conclusions}
\label{sec:conc}

In this paper, a parametric SCM for electrical resistance is proposed to study the effects of three experimental factors, using data collected on electronic devices operating under either an accelerated stress regime or a no-stress regime.
We believe that our proposal  represents the simplest non-trivial context from which several extensions are possible.
This is the reason why a comprehensive summary addressing all recent and ongoing methodological research is not provided here.
For example, \citet{correa2024} generalize Pearl's do-calculus from interventional to counterfactual reasoning   
and  introduce  a new graphical representation called Ancestral Multi-World Network. 
Nevertheless,  we consider only the factual world and a single counterfactual world in our context.

Model extensions are required when
different batches of devices are manufactured in separate production facilities, 
since transportability must be  considered in the analysis \citep{Bareinboim2016}.
This is also relevant  
if  both sources of information,  $\mathcal{D}_{AS,NS}$,
are available for the considered class of electronic devices.
Transferability makes it possible to learn about correspondent model parameters
across different  regimes or sources.

Future work might consider  more structured contexts, 
where interaction terms between investigated factors 
are needed.
The location parameter $\psi$ and the exponent $\tau$ might be
affected by epistemic uncertainty, therefore
additional observation times should be planned 
to contain the resulting uncertainty.
In our case study,   the observation time for each electronic device was also treated as  known without uncertainty,  
but in other contexts,
deviations from the nominal observation value might be of critical importance, e.g., 
very small  $\sigma^2_0, \sigma^2_{Y}$ and large deltas.

Similar considerations apply to the marginal independence of the exogenous random variables
 $\{(u_{0,1},\ldots,u_{0,n},\ldots, u_{t,1}, \ldots, u_{t,n}): t =1,2,3\}$.
According to our expert, this assumption is reasonable in the context 
presented here. 
However, it may not hold in more structured settings unless the model is appropriately extended.
In many contexts, model granularity is the key factor
that determines the extent to which the specified
model approximates the true data-generating process.

\subsection*{Funding}

The authors gratefully acknowledge the financial support from the
Italian Ministry of University and Research (MUR) 
under the Project of  Relevant National Interest (PRIN 2022),
project code E3DM - 
Experimental Design and Maintenance, a  Decision-Making approach driven by Degradation  Models - CUP master: B53C24006390006, CUP: B53C24006400006.
\begin{figure}[h!]
\center
\includegraphics[scale=2.00]{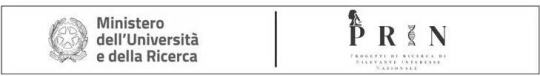} 
\end{figure}

\subsection*{Online supplements}
  
The project repository \textit{rel4dev} is currently under development, and the two datasets can be downloaded from it:
\url{https://github.com/federico-m-stefanini/rel4dev}

\clearpage
\newpage  
\bibliographystyle{agsm}

\end{document}